\documentclass[useAMS,usenatbib]{mn2e}

\title[Lense-Thirring precession]{Lense-Thirring precession around supermassive black holes during tidal disruption events}
\author[Franchini, Lodato and Facchini]{ Alessia Franchini$^{1}$\thanks{alessia.franchini@unimi.it}, Giuseppe Lodato$^{1}$\thanks{giuseppe.lodato@unimi.it} and Stefano Facchini$^2$\\
$^1$Dipartimento di Fisica, Universit\`a Degli Studi di Milano, Via Celoria, 16, Milano, 20133, Italy\\
$^2$Institute of Astronomy, Madingley Road, Cambridge CB3 OHA, UK
}
\usepackage{textcomp, libertine} 
\usepackage[english]{babel}
\usepackage[T1]{fontenc}
\usepackage[latin9]{inputenc}
\usepackage{float}
\usepackage{amsbsy}
\usepackage{amstext}
\usepackage{amssymb}
\usepackage{graphicx,txfonts}
\usepackage{esint}
\usepackage{epsfig}
\usepackage{xcolor,colortbl}
\usepackage[authoryear]{natbib}
\usepackage{url}

\begin{document}

\pagerange{\pageref{firstpage}--\pageref{lastpage}} \pubyear{2015}

\maketitle 

\label{firstpage}

\begin{abstract}
A tidal disruption event (TDE) occurs when a star wanders close enough to a black hole to be disrupted by its tidal force.
The debris of a tidally disrupted star are expected to form an accretion disc around the supermassive black hole. The light curves of these events sometimes show a quasi-periodic modulation of the flux that can be associated with the precession of the accretion disc due to the Lense-Thirring ("frame-dragging") effect. Since the initial star orbit is in general inclined with respect to the black hole spin, this misalignment combined with the Lense-Thirring effect leads to a warp in the disc. In this paper we provide a simple model of the system composed by a thick and narrow accretion disc surrounding a spinning supermassive black hole, with the aim to: (a) compute the expected precession period as a function of the system parameters, (b) discuss the conditions that have to be satisfied in order to have rigid precession, (c) investigate the alignment process, highlighting how different mechanisms play a role leading the disc and the black hole angular momenta into alignment.

\end{abstract}


\section{Introduction}\label{into}

Recently tidal disruption events (TDEs) have received significant attention because in principle they allow the indirect detection of quiescent black holes in the center of galaxies. In this way, not only we can probe the existence of dormant black holes but the modelling of single events can lead to constraints on the black hole parameters, such as its mass and spin value.

In general black hole masses can be measured through dynamical studies of the orbital motion of individual stars and gas in galactic nuclei. This kind of measurement can be performed using Newtonian mechanics since the stars are far from the black hole. The estimate of the spin is much more challenging. The spin has no gravitational effects in the Newtonian theory, so it can be measured only by probing the space-time geometry close to the black hole. Tidal disruption events occurring at distances of only a few gravitational radii from the black hole offer just this opportunity.
Tidal disruption events were discovered in the '90s as very bright, soft X-ray outbursts from otherwise quiescent galaxies by the \textit{ROSAT}  survey \citep{1999A&A...343..775K}.
Dedicated surveys at various wavelengths have been used to detect such signals. One of the first events in the optical, detected by \textit{Pan-STARRS} is known as PS1-10jh \citep{2012Natur.485..217G}. Previously, similar UV flares had been observed by the GALEX satellite \citep{2009ApJ...698.1367G}. 
Using SDSS (Sloan Digital Sky Survey) other two candidate TDEs have been identified in the optical band \citep{2011ApJ...741...73V}.
Other observations of TDE come from the \textit{XMM-Newton} survey \citep{2008A&A...489..543E}. Note that the X-ray emission traces the very vicinity of the supermassive black hole and provides an important probe of the central regions of galaxies.
The \textit{Swift} satellite led to a breakthrough in the field of TDEs. Its discovery of the first event that launched a relativistic jet, Swift J1644+57 \citep{2011Natur.476..425Z,2011Natur.476..421B,2011Sci...333..203B}, has triggered many theoretical studies on the formation of radio jets, and this event has now the best covered lightcurve of any TDE to date \citep{2015arXiv150501093K}. Subsequently, \cite{2012ApJ...753...77C} have discovered a second jetted TDE, Swift J2058.

The papers by \cite{1988Natur.333..523R}, \cite{1989IAUS..136..543P} and \cite{1989ApJ...346L..13E} have set the theoretical basis for the interpretation of such events.
A tidal disruption event occurs whenever a star wanders too close to a black hole, enough that the pericenter of its orbit $r_{\mathrm{p}}$ is of the order of the tidal radius 
\begin{equation}
r_{\mathrm{t}}=\left(\frac{M}{M_{*}}\right)^{1/3}R_{*}=0.47\,\mbox{AU}\,\left(\frac{M_{6}}{m_{*}}\right)^{1/3}x_{*}
\end{equation}
where  $M$ is the black hole mass, $M_{*}$ and $R_{*}$ are the stellar mass and radius respectively. We have also introduced the dimensionless quantities $M_{6}=M/(10^6\, M_{\odot})$, $x_{*}=R_{*}/R_{\odot}$ and $m_{*}=M_{*}/M_{\odot}$. We assume that the initial orbit of the star is parabolic. 

Usually, one defines the penetration factor $\beta=r_{\mathrm{t}}/r_{\mathrm{p}}$. Inside the tidal radius, thus for $\beta \gtrsim 1$, the tidal force of the black hole overcomes the internal self-gravity of the star. In this paper we assume $\beta=1$.

The stellar debris bound to the hole lie in highly eccentric orbits and at a time
\begin{equation}
t_{\mathrm{min}}\approx41\, M_{6}^{1/2}m_{*}^{-1}x_{*}^{3/2}\,\mbox{d\,}\label{eq:t_fb}
\end{equation}
after the tidal disruption they start coming back at pericenter at a rate \citep{2011MNRAS.410..359L}
\begin{equation}
\dot{M}_{\mathrm{fb}}=\dot{M}_{\mathrm{p}}\left(\frac{t}{t_{\mathrm{min}}}\right)^{-5/3}\label{eq:dot_m}
\end{equation}
where 
\begin{equation}
\dot{M}_{\mathrm{p}}=\frac{1}{3}\frac{M_{*}}{t_{\mathrm{min}}}\approx 1.9\times10^{26}\,M_{6}^{-1/2}m_{*}^{2}\,x_{*}^{-3/2}\mathrm{g\,s^{-1}}.\label{eq:dot_m_p}
\end{equation}
Comparing the peak value of Eq. (\ref{eq:dot_m_p}) with the Eddington limit
\begin{equation}
\dot{M}_{\mathrm{Edd}}= 1.3\,10^{24}M_{6}\left(\frac{\eta}{0.1}\right)^{-1}\mathrm{g\,s^{-1}},
\end{equation}
where $\eta$ is the accretion efficiency of the black hole, one can easily see that for a black hole with $M_{6}=1$, the fallback rate at peak can be as large as $100$ times the Eddington rate and is therefore
expected to produce a very bright flare. The ratio between the peak fallback accretion rate and the Eddington value scales as $\propto M^{-3/2}$, so the fallback rate is expected to be only marginally super-Eddington for a black hole with $M_{6}=10$.

Thus a distinguishing feature of tidal disruption events  is that, to a first approximation, the fallback rate $\dot{M}_{\mathrm{fb}}$ of the stellar debris onto the black hole should decrease with $t^{-5/3}$. If the fallback rate can be directly translated into an accretion luminosity ($L=\eta\dot{M}c^2$), one then expects the TDE luminosity to follow the same behaviour with time. Indeed, a $t^{-5/3}$ light curve is generally fitted to the observed luminosities of events interpreted as stellar disruptions \citep{2008ApJ...676..944G,2009ApJ...698.1367G,2012Natur.485..217G}.

In reality this power-law evolution is expected to occur at late times, while initially the light curve shows a smooth rise to the peak, the details of which depend on the internal structure of the disrupted star \citep{2009MNRAS.392..332L}. Also this behaviour has been sometimes observed \citep{2009ApJ...698.1367G,2012Natur.485..217G}.

The debris of the disrupted star that remain bound to the black hole are expected to form an accretion disc around the central object. The formation of the disc depends on the relative efficiency of three processes \citep{1989ApJ...346L..13E}: circularization, viscous accretion and radiative cooling. For non-spinning black holes, \cite{2013MNRAS.434..909H} and \cite{2015arXiv150104635B} have demonstrated through hydrodynamical simulations that the gas debris circularize on a few orbital timescale because relativistic precession causes the stream to self-cross, forming either a thin disc at the circularization radius $r_{\mathrm{c}}\sim 2r_{\mathrm{p}}$, or an extended thick torus depending on the cooling efficiency. \cite{2015arXiv150105306G} showed that in the case of spinning black hole the nodal precession can deflect material out of its original orbital plane such that an intersection is no longer guaranteed, resulting in a significant delay before circularization. Simulations performed by \cite{2015arXiv150105207H}  reveal that debris circularization depends sensitively on the efficiency of radiative cooling.
Recent numerical simulations by \cite{2015ApJ...804...85S} have shown that the circularization timescale is close to the period of the most bound initial orbit.
\cite{2015arXiv150205792P} suggest that is the energy liberated during the disc formation that powers the observed optical TDE candidates.

Since the accretion disc forms very close to the black hole, general relativity effects must be taken into account. One of these is the Lense-Thirring effect.
\cite{2012PhRvL.108f1302S} (SL) demonstrated that the Lense-Thirring precession around a spinning SMBH can produce significant time evolution of the disc angular momentum vector, since in general the initial orbit of the star is inclined with respect to the spinning black hole equatorial plane. 
For a thin disc it is expected that the Bardeen-Petterson effect \citep{1975ApJ...195L..65B} will induce a warp in the disc structure. \cite{2013ApJ...762...98L} argued that the precession of the jet is a possible consequence of this effect, providing an explanation of the Swift J1644 quasi-periodic modulation of the light curve.
However the disc formed after a TDE is expected to be geometrically thick rather than thin \citep{2009MNRAS.400.2070S,1999ApJ...514..180U}. In this configuration the Bardeen-Petterson effect does not occur. Still, if the disc is narrow it can precess as a solid body rotating around the black hole, inducing the precession of the jet and thus the modulation in the light curve.
Morover, the Lense-Thirring torque induces a warp in the disc. This warp can propagate in the diffusive regime if $\alpha\gtrsim H/R$ \citep{1992MNRAS.258..811P,2010MNRAS.405.1212L} or in the bending waves regime if $\alpha\lesssim H/R$ \citep{1999MNRAS.309..929N}, $\alpha $ being the \cite{1973A&A....24..337S} viscosity, and $H$ being the disc thickness.  Since the disc formed after a TDE is expected to be thick we are in the second regime, thus the warp propagates as a wave with half the speed of sound inside the disc.

\cite{2014ApJ...784...87S} have reexamined this issue considering a disc spreading both inwards and outwards from the circularization radius and its precession as a solid body applying the model to Swift J1644.

A precessing disc might offer a novel way to measure spin in black hole systems, through timing observations. The prospects for an effective determination of the spin parameter in this way are encouraged by the succesfull application of such techniques to stellar mass black holes, through the analysis of Type C QPOs in Low Mass X-ray Binaries (e.g.,  \citealt{Ingram2009,2014MNRAS.437.2554M}).

In this paper we consider a simple model with a disc that extends from the circularization radius (about twice the tidal radius) towards the innermost stable circular orbit around the supermassive black hole. Thus we develop the model proposed by \cite{2012PhRvL.108f1302S}, investigating the alignment process more in details.

The paper is organised as follows. In Section \ref{disc} we describe the structure of the accretion disc formed after a tidal disruption event.  In Section \ref{prec} we investigate the rigid precession of the disc due to the Lense-Thirring effect, compute the precession period for different values of the parameters of the system and investigate the conditions that have to be satisfied in order to have rigid precession. In Section \ref{time-ev} we describe our time dependent calculations and the first results, while Section \ref{alignment} contains the study of the alignment process and the calculation of the alignment timescale. Finally, in Section \ref{concl} we discuss our results and draw our conclusions.

\section{A simple model for the disc structure in tidal disruption events}\label{disc}

\begin{figure}
\includegraphics[width=0.5\textwidth]{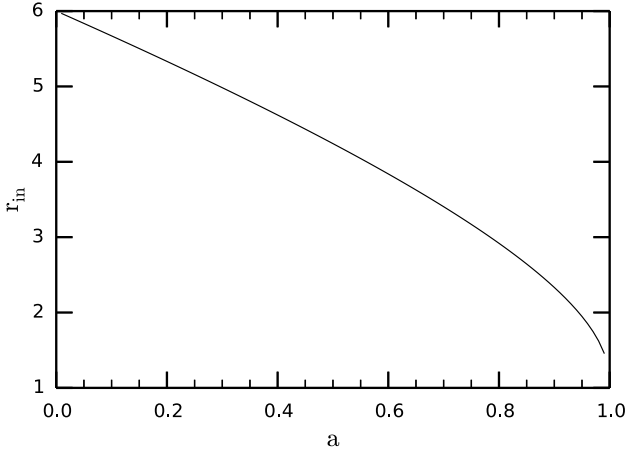}
\caption{\label{fig:a-vs-rin}Dimensionless inner radius of the disc as a function of the black hole spin $a$.}
\end{figure}

The accretion disc formed after the tidal disruption of a star by a supermassive black hole is expected to be narrow and thick \citep{1997astro.ph..8265U}.
The inner radius of the disc is
\begin{equation}
R_{\mathrm{in}}=r_{\mathrm{in}}\frac{GM}{c^{2}}=1.5\times10^{11}\, r_{\mathrm{in}}\, M_{6}\,\mbox{cm}\label{eq:rin}
\end{equation}
where $r_{\mathrm{in}}$ is the inner radius in units of the gravitational radius $R_{\mathrm{g}}=GM/c^{2}$. This corresponds to the innermost stable circular orbit (ISCO) for a black hole with spin $a$ and mass $M_{6}$ (Fig. \ref{fig:a-vs-rin} shows how the disc inner radius changes with the spin value). 

From now on we will use $r=R/R_{\mathrm{g}}$ as the dimensionless radial coordinate.
 
The spin $a$, or dimensionless angular momentum, is defined by the following expression for the black hole angular momentum
\begin{equation}
{J}_{\mathrm{h}}=a\frac{GM^2}{c},
\end{equation}
and its value is restricted $0<|a|<1$, where the lower limit refers to a Schwarzschild black hole for which no Lense-Thirring precession occurs.

We assume the outer radius of the disc to correspond to the circularization radius, that is about two times the tidal radius in the case $\beta=1$ :

\begin{equation}
R_{\mathrm{out}}=2\times0.47\,\mbox{AU}\,\left(\frac{M_{6}}{m_{*}}\right)^{1/3}x_{*}=1.41\times10^{13}M_{6}^{1/3}\,\mbox{cm\,}\label{eq:rout}
\end{equation}
where in the last equality we assumed that the disrupted star is a solar type one, thus $m_{*}=x_{*}=1$. 

In units of the gravitational radius the inner radius $r_{\mathrm{in}}$ is a function of the spin only, while the dimensionless outer radius depends only on the mass of the black hole, $r_{\mathrm{out}}=94\, M_{6}^{-2/3}$.

The disc extension is then given by 
\begin{equation}
\frac{r_{\mathrm{out}}}{r_{\mathrm{in}}}=94\,\frac{1}{r_{\mathrm{in}}}\, M_{6}^{-2/3}\label{eq:r_out}
\end{equation}
which takes into account the dependence of the disc extension on the black hole spin $a$ and on the black hole mass through $M_{6}$.

\begin{figure}
\begin{centering}
\includegraphics[width=0.5\textwidth]{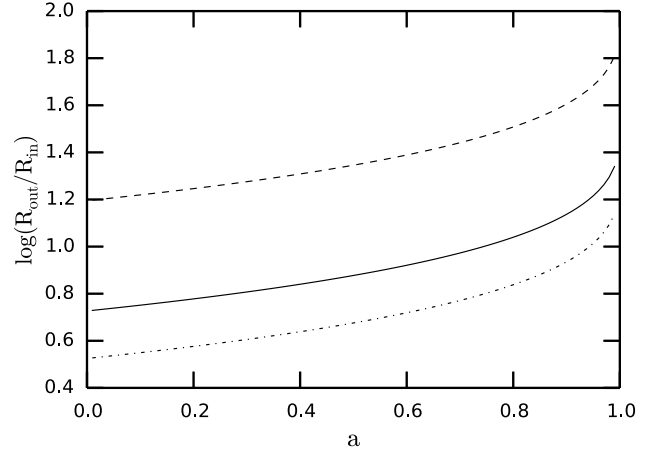}
\par\end{centering}

\caption{\label{fig:a-vs-rout/rin}Disc extension as a function of the black hole spin $a$ for different values of $M_{6}$.
The dashed line refers to $M_{6}=1$, the solid one to $M_{6}=5$ and the dash-dotted one to $M_{6}=10$. }
\end{figure}

Fig. \ref{fig:a-vs-rout/rin} shows the extension of the disc as a function of the spin for different values of the black hole mass.
From Fig. \ref{fig:a-vs-rin}, we can see that for larger values of $a$, $r_{\mathrm{in}}$ decreases and thus the disc inner radius is closer to the black hole, thus the disc is wider, since the outer radius does not depend on the spin of the black hole but only on its mass. 
We can see that the ratio $r_{\mathrm{out}}/r_{\mathrm{in}}$ in the most massive case of $10^{7}M_{\odot}$ is very small, from $3$ to roughly $10$ depending on the spin value. The disc is then relatively narrow. For lower black hole masses the ratio between the outer and the inner radius ranges from $\approx15$ to $\approx60$.

\subsection{A slim disc model for the disc structure}

Since the accretion disc is narrow as we have shown above and the accretion rate is close to the Eddington value, we can reasonably assume that the whole disc is radiation pressure dominated. This means that the sound speed is given by $c_{s}^2=\epsilon/3\rho$ where $\epsilon$ is the energy density of radiation and $\rho$ the gas density. 

Since the disc formed after a TDE is expected to be thick, we consider a slim disc model \citep{2009MNRAS.400.2070S,2012PhRvL.108f1302S} in order to describe the density and temperature profiles of the disc. Specifically we use the radial profile of the aspect ratio obtained by \cite{2009MNRAS.400.2070S}, who self-consistently solved the mass, momentum and energy conservation taking into account the advective heating term in the energy equation \citep[see also][]{Abramowicz1988}. Note that such a slim disc model has been often used in TDE modelling \citep{2009MNRAS.400.2070S,2012PhRvL.108f1302S,2011MNRAS.410..359L}. A proper modelling should involve global MHD, GR simulations, but for the simple analytical analysis proposed in this paper we deem this approximation as reasonable.

This approach leads to the following aspect ratio:
\begin{equation}
\frac{H}{R}=\frac{3}{2}(2\pi)^{1/2}\,\eta^{-1}\dot{m}\,r^{-1}f(r)\,K(r)^{-1}\label{eq:hoverr-gen}
\end{equation}
where $K$ is the following function of radius $r$ and spin $a$ \citep{2009MNRAS.400.2070S}:
\begin{equation}
K(r) = \frac{1}{2}+\left[\frac{1}{4} +6\,f(r)\left(\frac{\dot{m}_{\mathrm{fb}}}{\eta}\right)^2\left(\frac{1}{r}\right)^2\right]^{1/2}\,.
\end{equation}
This correction factor $K(r)$ takes into account the modified structure for a slim disc.
The sound speed is then given by:
\begin{equation}
c_{s} = \frac{3}{2}(2\pi)^{1/2}\,\eta^{-1}c\,\dot{m}_{\mathrm{fb}}\,r^{-3/2}f(r)\,K^{-1}(r).\label{c_s}
\end{equation}

The surface density dependency on the various quantities of interest is:
\begin{equation}
\Sigma=\Sigma_{0}r^{-3/5}f^{3/5}(r)\,\mathrm{g\,c{m^{-2}}}\label{eq:dens}
\end{equation}
where  $\Sigma_{0}$ is a constant that depends on $\dot{m}=\dot{M}/\dot{M}_{\mathrm{Edd}}$, $M_{6}$ and $\alpha$, and $f(r) = 1-(r_{\mathrm{in}}/r)^{1/2}$ .

In order to describe the evolution of the disc we consider a viscosity given by the $\alpha$-prescription by \cite{1973A&A....24..337S}. We assume the stress tensor to be proportional to gas pressure only rather than to total pressure, in order to avoid the \cite{1974ApJ...187L...1L} instability. In our model the $\alpha$-parameter is both constant in time and isotropic. This simple assumption is commonly taken in evolutionary models of discs in the literature, since it prevents computationally expensive simulations. In order to properly compute the viscous stress in discs simulations one would need to perform magnetohydrodynamic (MHD) calculations to treat the magneto-rotational instability (MRI) \citep[e.g.][]{Sorathia2013}. However the work presented here in this paper is a preliminary study with the aim to confirm the validity of the model, thus we consider the simple isotropic viscosity. 

\begin{figure*}
  \centerline{
  \includegraphics[width=0.5\textwidth]{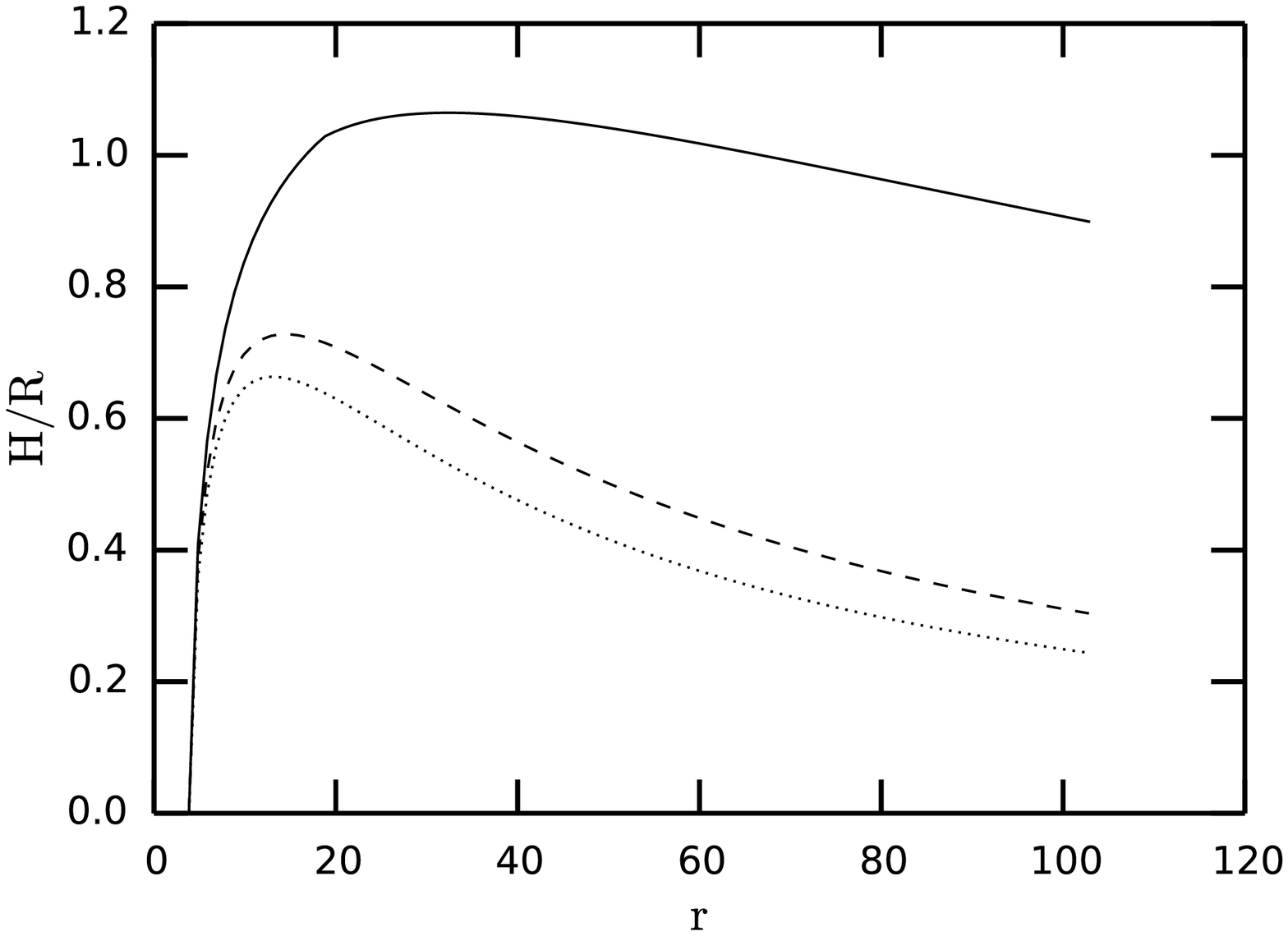}
  \includegraphics[width=0.5\textwidth]{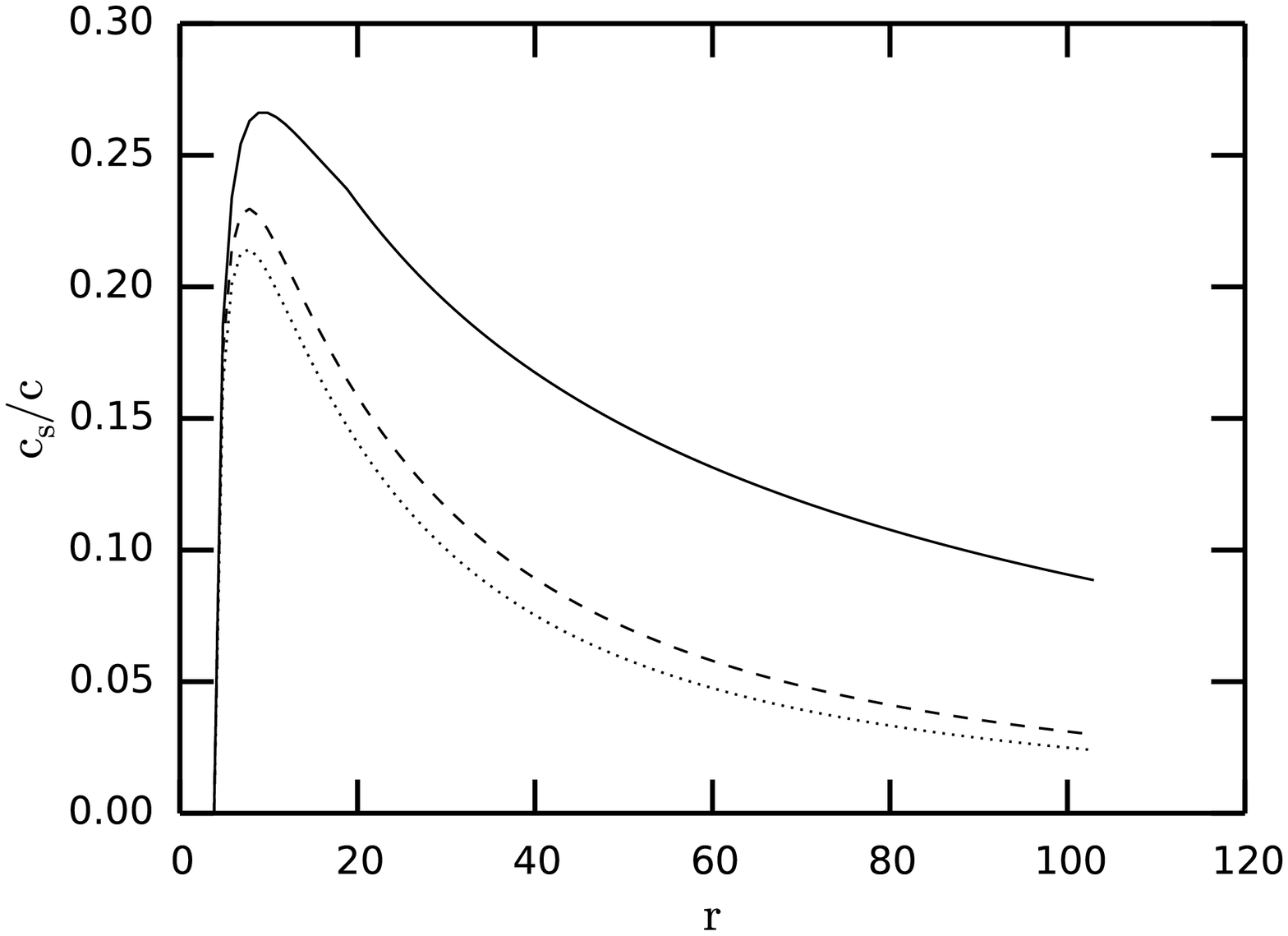}}
\caption{\label{fig:thickness-soundspeed}Disc ratio $H/R$ (left panel) and sound speed $c_{s}/c$ (right panel) as a function of the radius. The solid lines refer to the super-Eddington phase at a time $t_{\mathrm{min}}$. The dashed lines refer to the Eddington phase, reached after $2.6\,t_{\mathrm{min}}$ and the dotted lines refer to the sub-Eddington phase, after $3\,t_{\mathrm{min}}$. The curves refer to a black hole with $M_{6}=10$ and $a=0.6$.}
\end{figure*}

\subsection{Super-Eddington phase}

The viscous time in the disc is generally much shorter than the fallback time, which would then imply that the stellar debris are immediately processed through the disc, such that the black hole accretion rate equals the fallback rate \citep{1988Natur.333..523R}. In reality this might not happen for two different reasons. Firstly, disc accretion might start with a significant delay with respect to fallback. In this case most of the debris accumulate in a ring and only start accreting when fallback has faded significantly. In this case the accretion rate through the disc is dictated by viscous processes rather than by fallback \citep{cannizzo90,2014ApJ...784...87S}. Numerical simulations of the disc formation process do not yet clarify the extent to which such a delay occurs \citep{2015arXiv150104635B}. Here, we thus assume that -- if possible -- the accretion rate does coincide with the fallback rate. 
Even in this case, however, if the fallback rate is super-Eddington, most of the debris might be expelled from the system in the form of an outflow. The dynamics of super-Eddington discs is complex and not well understood. While some models seem to imply that super-Eddington rates can be sustained with little mass outflow if the disc is radiatively inefficient \citep{ohsuga07,McKinney2014}, it is in general expected that super-Eddington rates will result in an outflow. Here we make the simple and reasonable assumption that the accretion rate inside the disc $\dot{M}(R)$ varies such as to maintain the disc locally at the Eddington value ($L_{\mathrm{disc}}(R)= L_{\mathrm{Edd}}$) the excess mass being expelled in an outflow.
The local disc luminosity is $L_{\mathrm{disc}}(R) = GM\dot{M}(R)/2R$, while the Eddington limit is $L_{\mathrm{Edd}}=\eta\dot{M}_{\mathrm{Edd}}c^2$ and the efficiency can be written as $\eta=GM/(2R_{\mathrm{in}}c^2)$. 

At large radii, the local disc luminosity is below Eddington even if the fallback rate is globally super-Eddington. The local disc luminosity exceeds the Eddington value for radii smaller than  
\begin{equation}
r_{\mathrm{Edd}}=\dot{m}_{\mathrm{fb}}r_{\mathrm{in}},
\end{equation}
which is a function of time.  
We thus assume that $\dot{m}=\dot{m}_{\mathrm{fb}}$ for $r>r_{\mathrm{Edd}}$ and that the disc is locally Eddington limited for $r<r_{\mathrm{Edd}}$, resulting in
\begin{equation}
\dot{m}=\frac{r}{r_{\mathrm{in}}}.
\end{equation}

The evolution of the disc thickness for a typical choice of parameters as a function of time is shown in Fig. \ref{fig:thickness-soundspeed}. 
The solid line refers to the super-Eddington phase and one can see that $H/R \sim 1$ and thus the disc is thick. The ratio increases with radius until $r=r_{\mathrm{Edd}}$ while at larger radii it scales as $1/r$.
The long dashed line refers to the Eddington phase while the dotted line refers to the sub-Eddington phase. 
The time at which the accretion rate is equal to the Eddington value can be inferred by requiring that $\dot{m}_{\mathrm{fb}}=1$. This gives 
\begin{equation}
t_{\mathrm{Edd}}=\left(\frac{\dot{M}_{\mathrm{Edd}}}{\dot{M}_{\mathrm{p}}}\right)^{-3/5}t_{\mathrm{min}}\label{eq:t-edd}.
\end{equation}
In Fig. \ref{fig:thickness-soundspeed} the three curves are obtained for a $10^7M_{\odot}$ black hole, spinning with $a=0.6$. For higher values of the black hole spin, the efficiency $\eta$ increases slightly and thus the disc thickness is lower.

\section{Global precession}\label{prec}

In the previous section we have described the structure of an accretion disc formed after the tidal disruption of a star by a supermassive black hole. Since the initial stellar orbit is likely to be inclined with respect to the black hole equatorial plane, the disc angular momentum is in general misaligned with the black hole spin. We now focus on the evolution of such inclined disc. 

The misalignment between the disc and the black hole spin generates a relativistic torque on the disc (Lense-Thirring effect). Such torque is proportional to $R^{-3}$, as we will see below. A faster precession implies an enhanced viscous dissipation, thus the inner disc tends to align with the equatorial plane while the outer regions keep the original misalignment. Additionally, under certain conditions the warped disc is expected to rigidly precess around the hole \citep[e.g.][see Section \ref{condition}]{1995MNRAS.274..987P,Fragile2005,Fragile2007}.   Observationally, the only evidence we have of this phenomenon might be the early periodicity of the light curve of the event Swift J1644+57 \citep[e.g.][]{2014ApJ...784...87S}. Here we aim to compute the precession period of a misaligned disc as a function of the black hole mass and spin, such that the observed periodicity in the light curve of TDEs may be used to infer fundamental properties of the central black hole.

\subsection{Precession period}\label{period}

In this section we assume that the disc can rigidly precess. We will discuss the conditions for rigid precession in Section \ref{condition}. We estimate the global precession frequency using the same procedure as in \citet{2013MNRAS.433.2157L}. At a given radius $R$ within the disc, the local external torque density is $\mathbf{T}(R)=\mathbf{\boldsymbol{\Omega}}_{\mathrm{LT}}(R)\times\mathbf{L}(R)$ \citep[e.g.][]{2006MNRAS.368.1196L,2013MNRAS.433.2157L}, where $\mathbf{\boldsymbol{\Omega}}_{\mathrm{LT}}(R)$ is the Lense-Thirring local precession frequency induced by the misalignment and $\mathbf{L}(R)=\Sigma\Omega R^2 \mathbf{l}$ is the angular momentum of the disc per unit area. 
We define the vector $\mathbf{\boldsymbol{\Omega}}_{\mathrm{LT}}(R)$, and thus the black hole spin, to be along the $z$-direction. If the misalignment is small, we can write the scalar equation for the local external torque density
\begin{equation}
T(R)=\Omega_{\mathrm{LT}}(R)L(R)\label{eq:torque-local},
\end{equation}
where $L(R)=\Sigma R^{2}\Omega\sqrt{l_{x}^{2}+l_{y}^{2}}\,$  is the angular momentum projected on the $xy$ plane. The variable $\Omega_{\mathrm{LT}}(R)$
is the precessional frequency at which an isolated ring of material would precess under the influence of the external torque $T(R)$.

Now we assume that the whole disc precesses with the same frequency $\Omega_{\mathrm{p}}$ and then write 

\begin{equation}
T_{\mathrm{tot}}=\Omega_{\mathrm{p}}L_{\mathrm{tot}},
\end{equation}
where we have defined the two quantities

\begin{equation}
T_{\mathrm{tot}}=\int_{R_{\mathrm{in}}}^{R_{\mathrm{out}}}\Omega_{\mathrm{LT}}(R)L(R)2\pi R\, dR
\end{equation}
and 
\begin{equation}
L_{\mathrm{tot}}=\int_{R_{\mathrm{in}}}^{R_{\mathrm{out}}}L(R)2\pi R\, dR.
\end{equation}
These are appropriate for small amplitude warps, which is the case we consider here. If we assume that the disc precesses as a solid body, we obtain the following relation for $\Omega_{\mathrm{p}}$:

\begin{equation}
\Omega_{\mathrm{p}}=\frac{\int_{R_{\mathrm{in}}}^{R_{\mathrm{out}}}\Omega_{\mathrm{LT}}(R)L(R)2\pi R\, dR}{\int_{R_{\mathrm{in}}}^{R_{\mathrm{out}}}L(R)2\pi R\, dR}.\label{eq:freqP}
\end{equation}
Having defined an expression for $\Sigma(R)$ with Eq. (\ref{eq:dens}), we now need to derive an expression for $\Omega_{\rm LT}(R)$, in order to compute the precession frequency. We evaluate the external torque using the expression:
\begin{equation}
\mathbf{T}=-\Sigma R^{2}\Omega\left(\frac{\Omega_{z}^{2}-\Omega^{2}}{\Omega^{2}}\right)\frac{\Omega}{2}\mathbf{e_{z}\times l}\label{eq:torque-vect},
\end{equation}
where $\mathbf{e_{z}}$ is the unit vector parallel to the black hole spin and $\Omega_{z}$ is the vertical oscillation frequency due to frame dragging. The approximate expressions for the angular and  vertical frequencies are \citep{1990PASJ...42...99K,2002MNRAS.337..706L}:

\begin{equation}
\Omega=\frac{c^{3}}{GM}\frac{1}{r^{3/2}+a}\label{eq:omega};
\end{equation}

\begin{figure}
\includegraphics[width=0.5\textwidth]{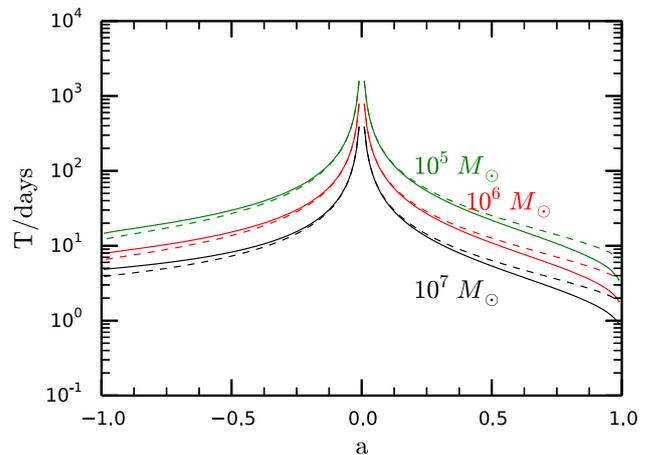}
\caption{\label{fig:glob_freq}Dependence of the global precession period on $a$ for a prograde and a retrograde disc. The dashed lines show the same period using the first order approximated forms eq. (\ref{eq:omega-approx}) and (\ref{eq:omegaLT-approx}). This results refer to $M_{6}=0.1$ (green lines), $M_{6}=1$ (red lines) and $M_{6}=10$ (black lines).}
\end{figure}

\begin{equation}
\left(\frac{\Omega^{2}-\Omega_{z}^{2}}{\Omega^{2}}\right)=4ar^{-3/2}-3a^{2}r^{-2}\label{eq:omega_z},
\end{equation}
where $-1<a<1$ is the dimensionless hole angular momentum ($a<0$ for a retrograde disc).

For $r \gg1$, we have 
\begin{equation}
\Omega=\frac{c^{3}}{GM}\frac{1}{r^{3/2}}\label{eq:omega-approx},
\end{equation}

From (\ref{eq:omega}) and (\ref{eq:omega_z}), we obtain the following expression for the local precession frequency, in units of the gravitational radius:

\begin{equation}
\Omega_{\mathrm{LT}}(r)=\frac{1}{2}\frac{c^{3}}{GM}\frac{1}{r^{3/2}+a}(4ar^{-3/2}-3a^{2}r^{-2}),\label{eq:local-prec}
\end{equation}
which, for $r\gg1$, gives the usual $r^{-3}$ dependence:

\begin{equation}
\Omega_{\mathrm{LT}} = 2a\frac{c^3}{GM}r^{-3}\,.\label{eq:omegaLT-approx}
\end{equation} 

Fig. \ref{fig:glob_freq} shows the dependence of the global precession period $T=2\pi/\Omega_{\mathrm{p}}$
on the black hole spin for three different values of the hole mass. The solid lines represent the estimate of the precession period obtained from numerical integration of (\ref{eq:freqP}), including all the terms of the expressions of the Keplerian
and vertical frequency (\ref{eq:omega}) and (\ref{eq:omega_z}). The dashed lines indicate
the same result where we have considered only the dominant term in the limit of $r\gg1$ in the same equations (Eq. \ref{eq:omega-approx} and \ref{eq:omegaLT-approx}). The difference between the two cases becomes relevant at high spin values, where the difference in the precession period can be as large as a factor of $2$, whereas it is negligible at low spin values. 

If the black hole has a large spin, the disc is wider than for low spin values (see Fig. \ref{fig:a-vs-rout/rin}) and has therefore a larger angular momentum. However, since the inner radius is closer to the hole, the torque that generates the precession is stronger. The combined effect is that the precession frequency $\Omega_{\mathrm{p}}$ is larger. In this case, the period is short enough to be detectable at the beginning of a TDE ($\sim1-20$\,days). 

\subsection{Condition for rigid precession}\label{condition}

So far we have determined the precessional frequency assuming that the disc does globally precess around the black hole. However, simulations and analytic studies have shown that often the disc does not precess as a solid body. One alternative possibility is that the disc can tear apart in discrete rings, which then precess with their local precession frequencies \citep[e.g.][]{1996MNRAS.282..597L,2010MNRAS.405.1212L,2012ApJ...757L..24N,2013MNRAS.434.1946N,2013MNRAS.433.2142F,2015MNRAS.449.1251D,2015MNRAS.448.1526N}. \citet{2013MNRAS.434.1946N} have suggested that such mechanism occurs in discs with a large misalignment, when the local precession is faster than the wave communication. When the same condition on the timescales is verified, but the misalignment is more moderate, the disc will tend to get aligned to the plane perpendicular to the spin of the black hole. \citet{KrolikHawley2015} showed through simulations that a time-steady transition can be achieved between an inner disk region aligned with the equatorial plane and an outer region orbiting in a different plane. Thus in this case there is no tearing.

\citet{1995MNRAS.274..987P}, \citet{1996MNRAS.282..597L} and \citet{Fragile2005} have shown that the most stringent condition for a disc to globally precess is that the local precession period at any radius ($t_{\rm p}=2\pi/\Omega_{\rm LT}$) is longer than the sound crossing time $t_{\rm wave}$.
In our case, this condition requires the local precession period at the inner edge to be longer than the time it takes a sound wave to communicate the torque to the whole disc. If such communication is inefficient, the disc either tears apart, or dissipates the internal stresses, thus going towards an aligned state.

The time dependent equations discussed below in Section \ref{time-ev} contain all the physics that regulates whether a disc can rigidly precess. We have tested the analytic condition with such equations for a disc with a simple power law for the sound speed and the surface density profiles. We obtain a good agreement between the numerical simulations and the analytic prediction, indicating that the warp evolution equations indeed contain the physics regulating the global precession condition. We are therefore confident in using the code itself to track the evolution of the radiation pressure dominated discs of this paper, and check {\it a posteriori} whether the disc globally precesses. Such method is simpler than verifying the analytic condition of all the simulated discs during the whole evolution (the properties of the disc vary with time), given the non trivial sound speed and surface density profiles (see equations \ref{eq:dens}-\ref{c_s}).

In general, we anticipate that global disc precession is always observed, initially at least, in all our simulations across the whole parameter space.

\section{Time dependent calculations}\label{time-ev}

In this section we consider the evolution of the disc with time to check that rigid precession does occur and how it evolves with time.
The accretion rate is super-Eddington at the beginning and decreases as $t^{-5/3}$.
This decrease with time implies a decrease of the ratio $H/R$ (see eq. (\ref{eq:hoverr-gen})). When $H/R$ drops below the value of the disc viscosity $\alpha$, the warp propagates inside the disc in a diffusive way \citep[][]{1983MNRAS.202.1181P,1992MNRAS.258..811P,2010MNRAS.405.1212L}. At this point the accretion flow is susceptible to the Bardeen-Petterson effect \citep{1975ApJ...195L..65B} that leads to alignment between the disc and the hole angular momentum.

\begin{figure*}
  \centerline{
  \includegraphics[width=0.5\textwidth]{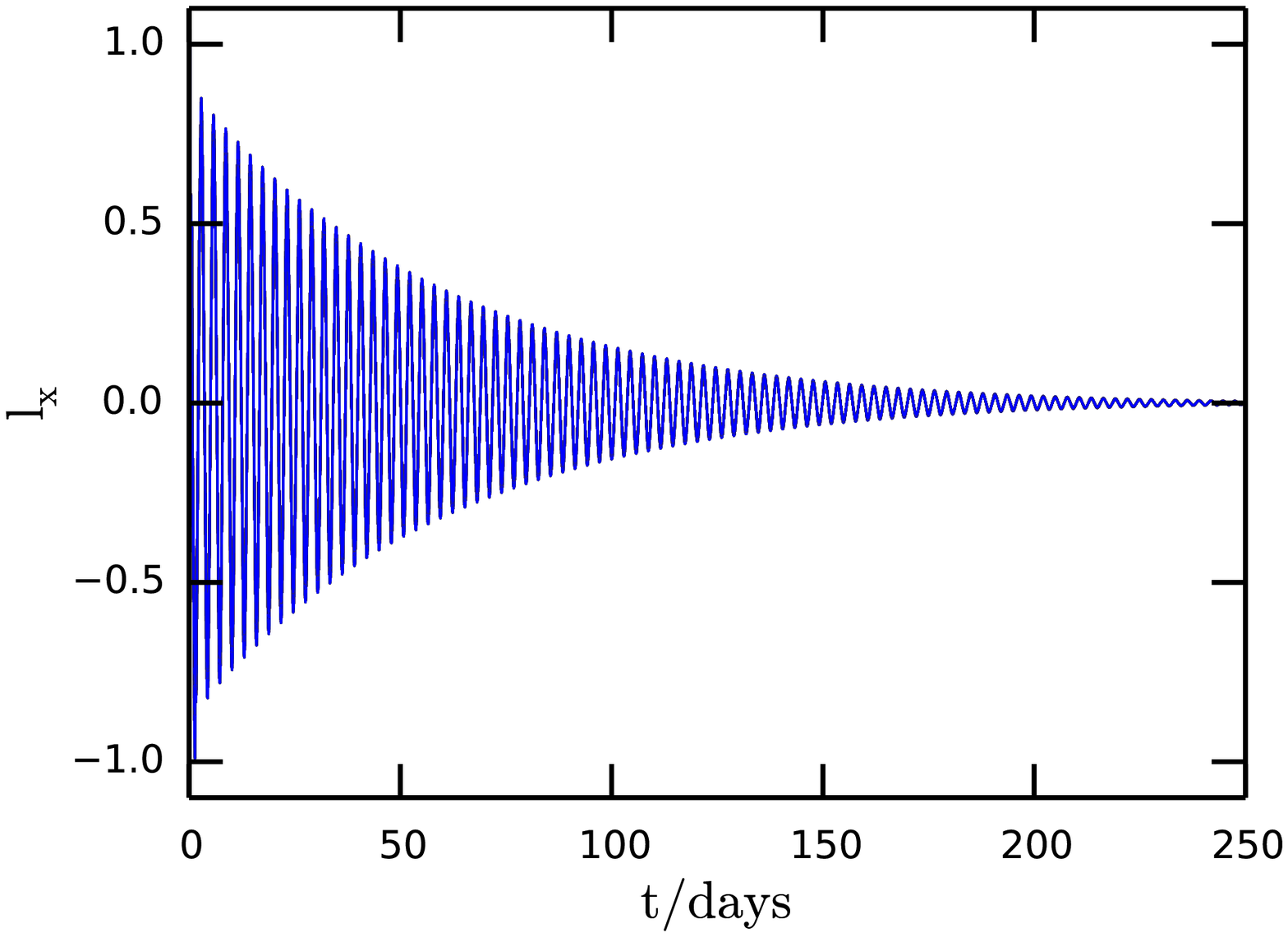}
  \includegraphics[width=0.5\textwidth]{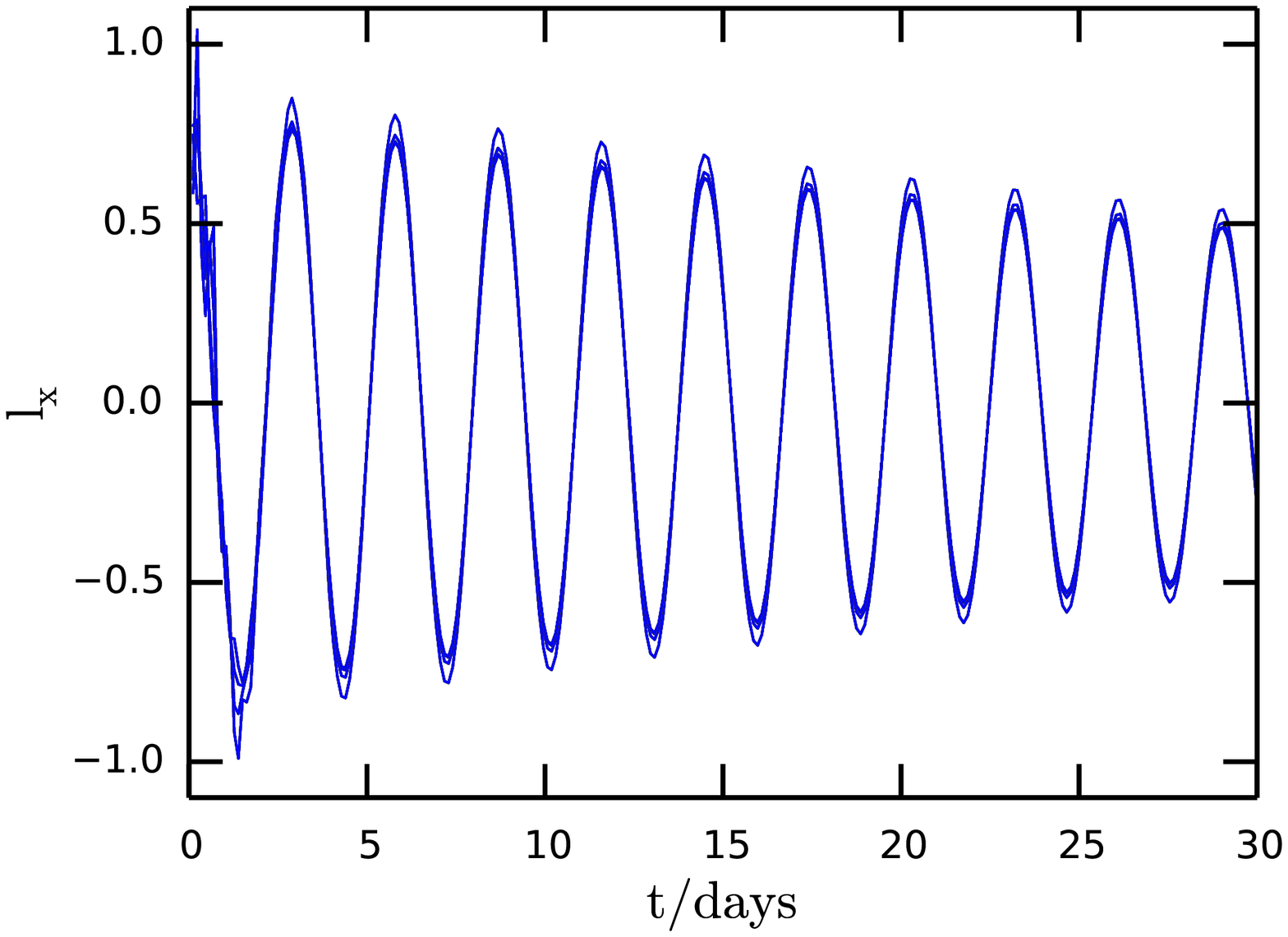}}
\caption{Evolution of the $x$-component of the disc angular momentum with time. The viscosity is $\alpha=0.05$, the value of the hole spin is $a=0.7$ and the mass $10^7M_{\odot}$. The right panel is a detail of the left one. The different curves refer to different radii inside the disc. Although the amplitudes are slightly different, the period of the oscillations at different radii is the same.}
\label{fig:prec-0.05-a0.7}
\end{figure*}

\begin{figure*}
  \centerline{
  \includegraphics[width=0.5\textwidth]{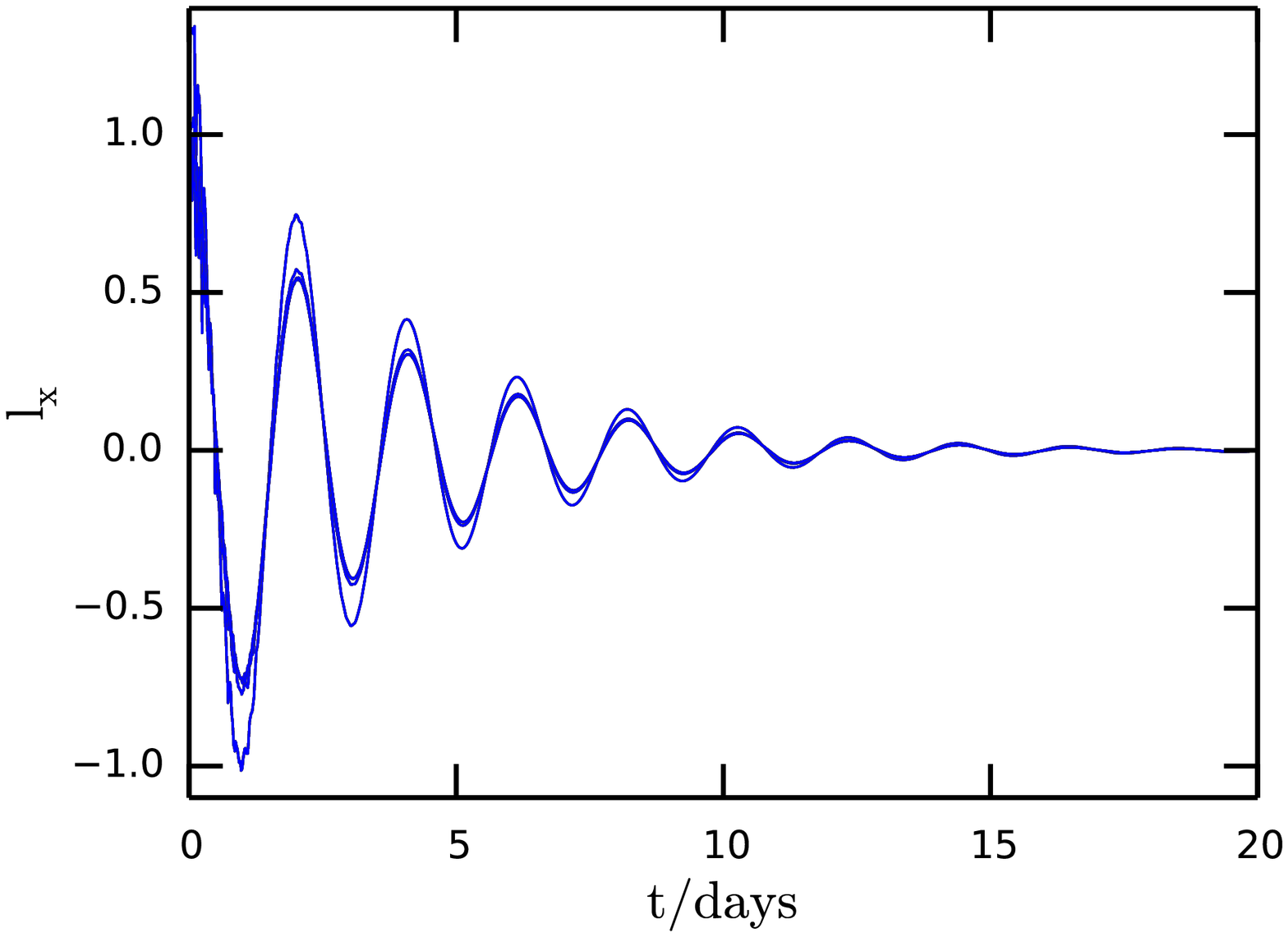}
  \includegraphics[width=0.5\textwidth]{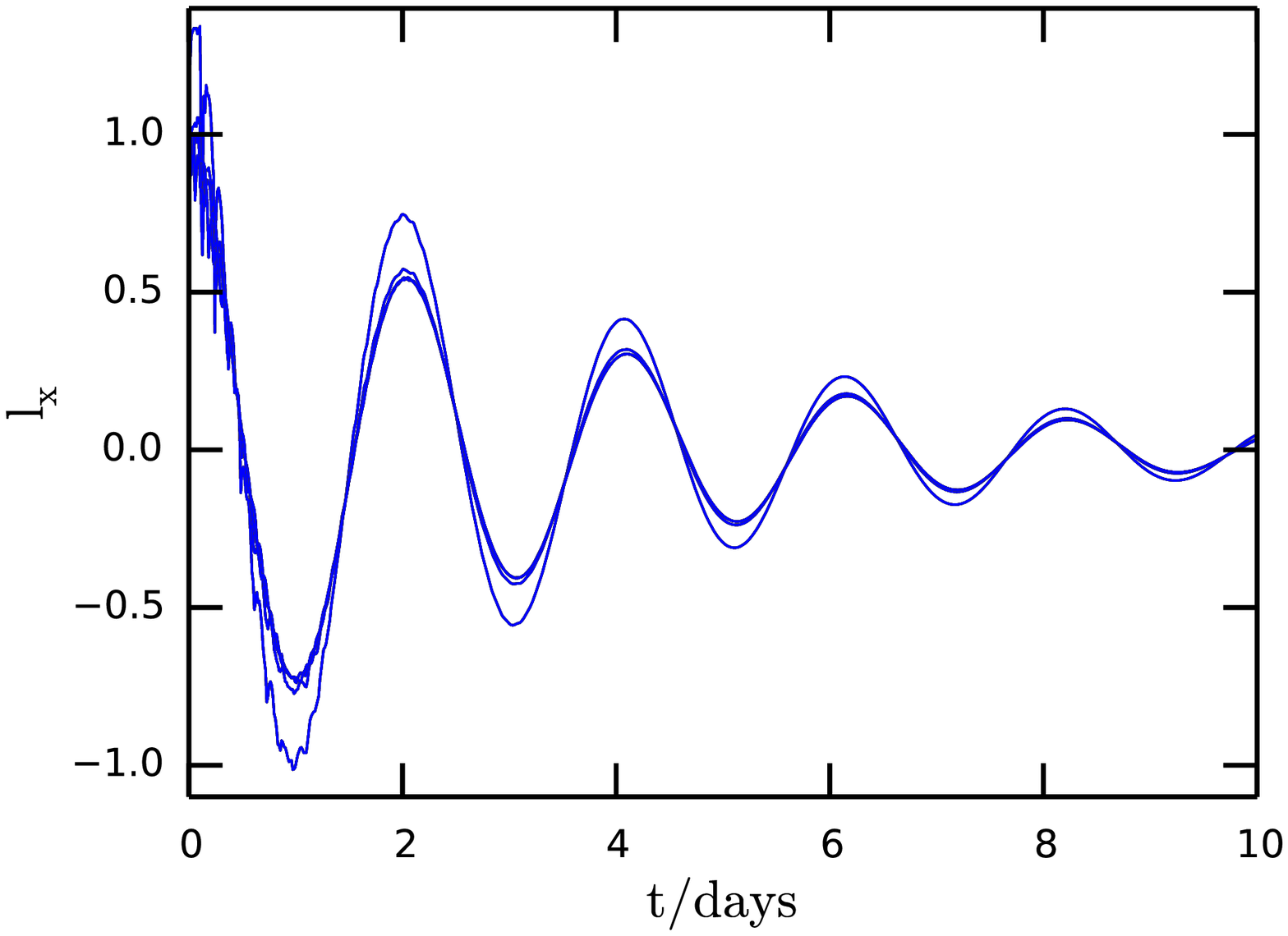}}
\caption{Evolution of the $x$-component of the disc angular momentum with time. The viscosity is $\alpha=0.05$, the value of the hole spin is $a=0.9$ and the mass $10^6M_{\odot}$. The right panel is a detail of the left one. The different curves refer to different radii inside the disc. Although the amplitudes are slightly different, the period of the oscillations at different radii is the same.}
\label{fig:prec-0.05-a0.9}
\end{figure*}

\subsection{Warp propagation in bending waves regime}
The disc formed after a tidal disruption event is expected to be thick (see upper panel in Fig. \ref{fig:thickness-soundspeed}) enough that $H/R\gtrsim\alpha$ and the warp caused by the Lense-Thirring effect propagates in the bending waves regime \citep{1983MNRAS.202.1181P}. Its propagation is described by the following linearised equations \citep{1995ApJ...438..841P,2000ApJ...538..326L}:

\begin{equation}
\Sigma R^{3}\Omega\frac{\partial\mathbf{l}}{\partial t}=\frac{\partial\mathbf{G}}{\partial R}+\mathbf{T}\label{eq:bw1}
\end{equation}
\begin{equation}
\frac{\partial\mathbf{G}}{\partial t}+\left(\frac{\kappa^{2}-\Omega^{2}}{\Omega^{2}}\right)\frac{\Omega}{2}\mathbf{e_{z}}\times\mathbf{G}+\alpha\Omega\mathbf{G}=\Sigma R^{3}\Omega\frac{c_{\rm s}^{2}}{4}\frac{\partial\mathbf{l}}{\partial R}\label{eq:bw2}
\end{equation}
where $\mathbf{T}(R,t)$ is the external torque density acting on the disc, that arises from a lack of spherical symmetry in the potential, and $\mathbf{G}(R,t)$ is the internal torque. Equation (\ref{eq:bw1}) represents the horizontal component of the angular momentum conservation equation, including the external torque.
The term on the right side of equation (\ref{eq:bw2}) depends on the speed of sound $c_{\rm s}$ which changes with time since it depends on $\dot{m}(t)$, see Eq. \ref{c_s}.
Both equation are valid if one requires the deviation from the Keplerian potential to be small, i.e., $|\kappa^2-\Omega^2|\lesssim \delta\Omega^2$ and $|\Omega_{z}^2-\Omega^2|\lesssim \delta\Omega^2$ where $\delta=H/R$ \citep{2014MNRAS.445.1731F}.

Note that the above equations, being linear, do not treat the possibility of shocks in the disc. The small warps produced in the case considered here (see below) do not produce supersonic shear and thus are not expected to develop shocks. During the disc formation phase \citep{2015ApJ...804...85S,2015arXiv150104635B} shocks might develop. However, here we treat the disc dynamics after the initial circularization phase.

The equations were derived in the case of isotropic viscosity $\nu=\alpha c_{\rm s}H$. 
\cite{1992MNRAS.258..811P} pointed out that the viscosity may be significantly anisotropic due to the different types of shear present in a warped disc. In general the viscosity parameter $\alpha$ in eq. (\ref{eq:bw2}) can be different from that responsible for the radial transfer of mass and angular momentum. 
Azimuthal shear is secular where gas particles drift further apart, whereas vertical shear is oscillatory and thus should induce less dissipation.
However, if the velocity spectrum of the turbulence is predominantly on scales $<H$ \citep{2012MNRAS.422.2685S} then it is
likely to act similarly in each direction. 
Recently, \citet{2015MNRAS.450.2459N} explored models for a disc that is tilted and forced to precess by the radiation warping instability, to see what constraints can be provided on the internal communication of angular momentum in a warped disc.

We used a numerical grid code in order to solve the above equations in both space and time domain.
In our code we used the radius $R$ as the only spatial variable of the system. We therefore discretized the disc into a set of thin annuli, each of which can be tilted and interacts with the others via pressure and viscous forces. 
We wrote the bending waves equations in terms of dimensionless quantities and then we solved them using the leapfrog algorithm.  
This code is an modified version of the code used by \citet{2013MNRAS.433.2157L}. In the linear regime it has proved to recover full 3D SPH simulations \citep[e.g.][]{2013MNRAS.433.2142F,2015MNRAS.448.1526N}.
 
\subsection{Results}

We performed simulations choosing the values of the black hole parameters mass and spin referring to the observed period of the event Swift J1644+57. This tidal disruption event has an observed period of $T=2.7$ days and the quasi-periodicity of the light curve lasts approximatively $10$ days \citep{2013ApJ...762...98L}. 
Based on Fig. \ref{fig:glob_freq}, we see that a period of $2.7$ d can be reproduced with $a=0.7$ for $M=10^7M_{\odot}$ and $a=0.9$ for $M=10^6M_{\odot}$. We have performed simulations also for $M<10^6M_{\odot}$.

The evolution of the disc with time is shown in Figs.  \ref{fig:prec-0.05-a0.7} and \ref{fig:prec-0.05-a0.9} for the two masses, respectively. The four curves refer to four different radii inside the disc at which the angular momentum is evaluated. As one can easily see, the amplitude changes slightly with radius but the curves have exactly the same period, implying that rigid precession is indeed occurring.
The values of the rigid precession period computed numerically agree well with the expected value, $2.7$ d.

As one can see (Figs.  \ref{fig:prec-0.05-a0.7} and \ref{fig:prec-0.05-a0.9}) the amplitude of the oscillation decreases steadly with time and the disc eventually aligns with the black hole. The most notable difference between the two cases shown is that while for $M=10^7M_{\odot}$ alignment occurs after several months, for $M=10^6M_{\odot}$ it only takes a few days to align for the same value of $\alpha$. We will discuss the alignment process in details below. 

We conclude that if we want to apply this model to Swift J1644, where some periodicity is observed only within the first $10$ days, we would disfavour a high SMBH mass, unless $\alpha$ is significantly larger than $0.05$. Similarly a value of $M$ below $10^6M_{\odot}$ is disfavoured since it would require an almost maximally spinning black hole. In particular, a $2.7$ d period can be reproduced only for an extremely high value of $a$ if $M<5\,10^5M_{\odot}$ (see the green lines in Fig. \ref{fig:glob_freq}).

\begin{figure}
\includegraphics[width=0.5\textwidth]{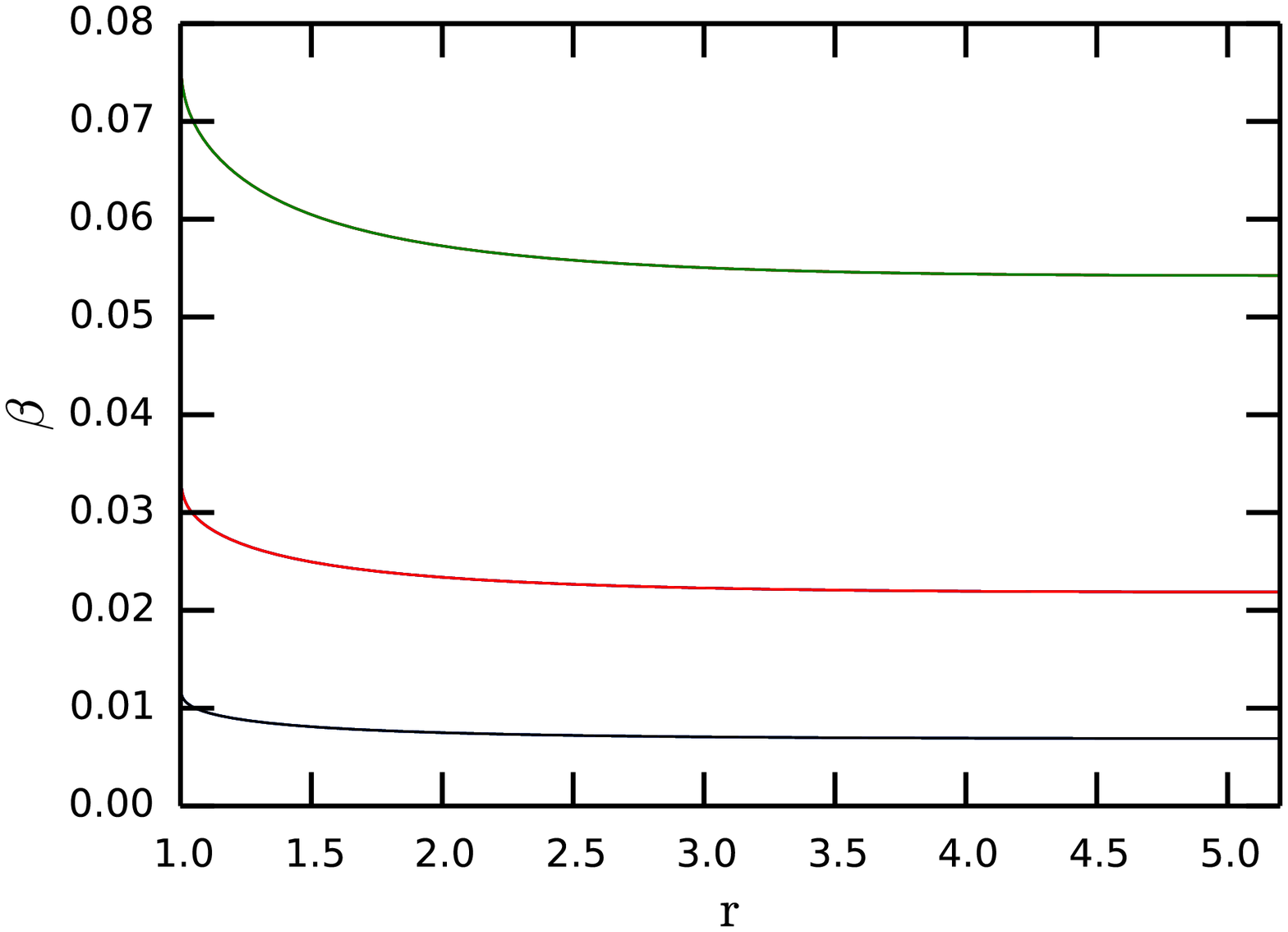}
\includegraphics[width=0.5\textwidth]{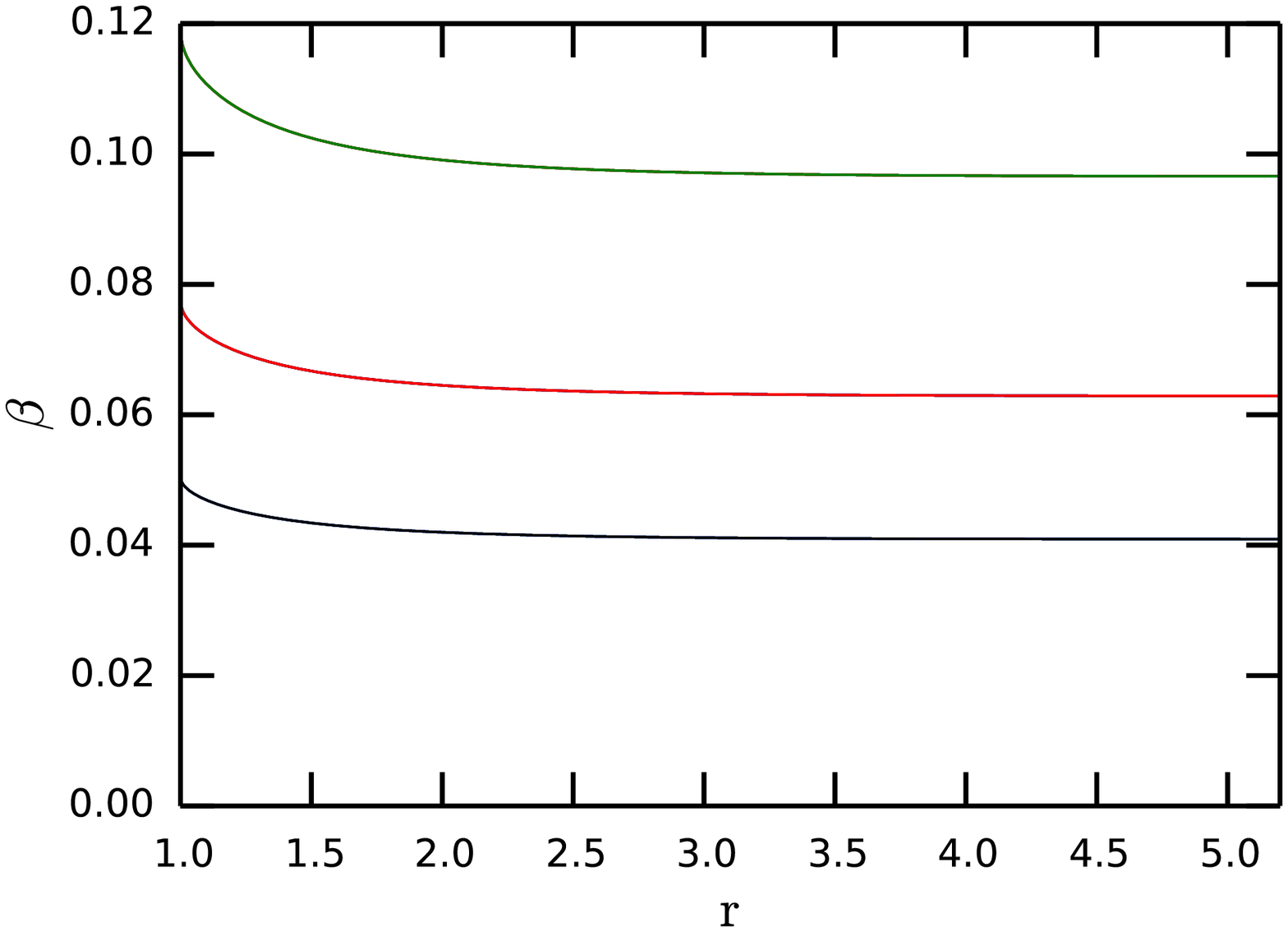}

\caption{\label{fig:tilt-alpha} Shape of the tilt $\beta$ inside the disc as a function of radius. Both panel refers to a black hole with $M=10^6M_{\odot}$ and $a=0.6$. The top panel was obtained taking $\alpha=0.05$. The green, red and black lines refer to the shape after roughly $290,\,350$ and $405$ days. The bottom panel was obtained with $\alpha=0.5$. The green, black, red and black lines refer to roughly $29,\,35$ and $40.5$ days.}
\end{figure}

Fig. \ref{fig:tilt-alpha} shows the tilt $\beta$ (defined as the misalignment between the spin and the local disc angular momentum) as a function of the radius for the case $M=10^6M_{\odot}$, $a=0.6$ and $\alpha=0.05$ (top) and $0.5$ (bottom), respectively. The first thing to notice is that while the disc is rigidly precessing, it maintains a small warp ($\beta$ is a function of the radius), which is necessary to balance the external torque. Note that the disc turns out to be (slightly) more misaligned in the inner regions than in the outer regions. This is not unexpected. As shown by \citet{2002MNRAS.337..706L} (see also more recently \citealt{2013MNRAS.433.2142F}) the radial profile of beta depends on the relative signs of the nodal and apsidal precession frequencies. While, for example, in the case of a circumbinary disc, the tidal torque produces a monotonically increasing tilt, in the case of Lense-Thirring precession the tilt function is not monotonic and can show regions where the disc warps away from alignment.

On a timescale much longer than the precession time, the disc slowly aligns. Comparing the two panels of Fig. \ref{fig:tilt-alpha} we note that the alignment time scales inversely with $\alpha$ and is thus proportional to the viscous timescale.

\section{Alignment}\label{alignment}

In general the mechanisms that are likely to lead to alignment are at least two. The first one can be the cooling of the disc, as a consequence of the reduction of the accretion rate (given by eq. \ref{eq:dot_m}). As the disc cools, it naturally becomes thinner.  Eventually the disc will transit to a regime where the warp propagation becomes diffusive ($\alpha \geq H/R$). In this regime there is no rigid precession and the disc rapidly aligns with the black hole spin. This is the case considered by \cite{2012PhRvL.108f1302S} (SL), who calculate the associated alignment timescale as
\begin{equation}
t_{\mathrm{thin}}=t_{\mathrm{min}}\alpha^{-3/5}\left[5\,f(r)\frac{M_{*}}{\dot{M}_{\mathrm{Edd}}t_{\mathrm{min}}}\frac{1}{r}\right]^{3/5}\label{eq:t-thin}
\end{equation} 
where $t_{\mathrm{min}}$ is the time after which the debris return to the pericenter (eq. \ref{eq:t_fb}). The alignment timescale in this case is $\propto \alpha^{-3/5}$, while we have just seen (Fig. \ref{fig:tilt-alpha}) that at least for the case $M=10^6M_{\odot}$, $a=0.6$ our results indicate that $t_{\mathrm{align}}\propto\alpha^{-1}$.

The second possible mechanism is provided by the presence of the natural disc viscosity that damps the oscillations at a rate inversely proportional to the viscosity ($t_{\mathrm{align}}\propto \alpha^{-1}$). Thus if the viscosity is higher, the warp is dissipated on a shorter timescale.

We investigate under which condition one process dominates over the other by running several simulations with different values of $\alpha$ and of the black hole parameters.
In general one may expect that since for low spin values the inner disc radius is larger, the torque exerted on the disc is weaker and rigid precession is expected to last on a longer timescale. 
The mass of the black hole changes the disc radial extent, thus we investigate for each values of the spin both values $10^6M_{\odot}$ and $10^7M_{\odot}$.

\begin{figure}
\includegraphics[width=0.5\textwidth]{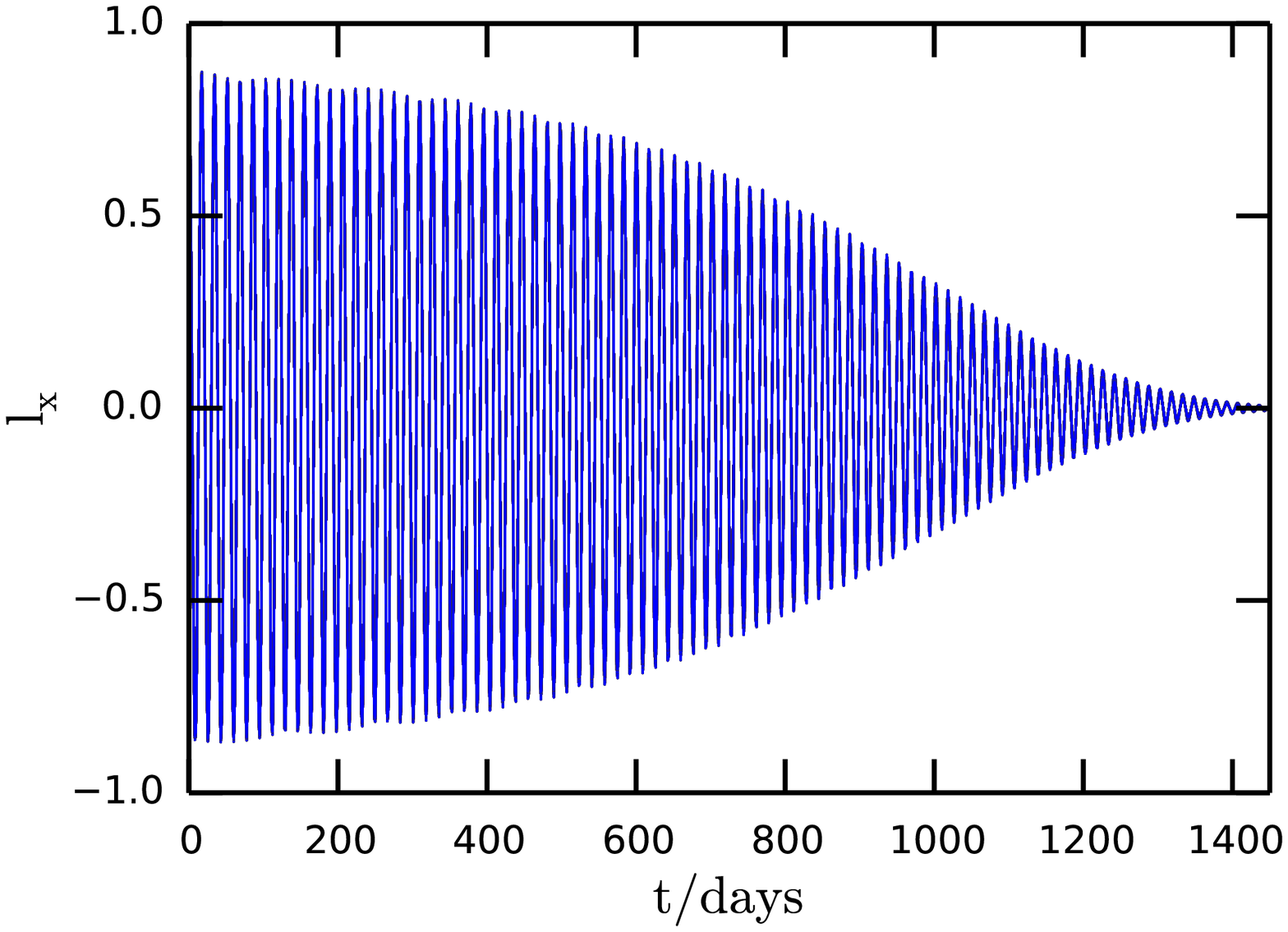}
\includegraphics[width=0.5\textwidth]{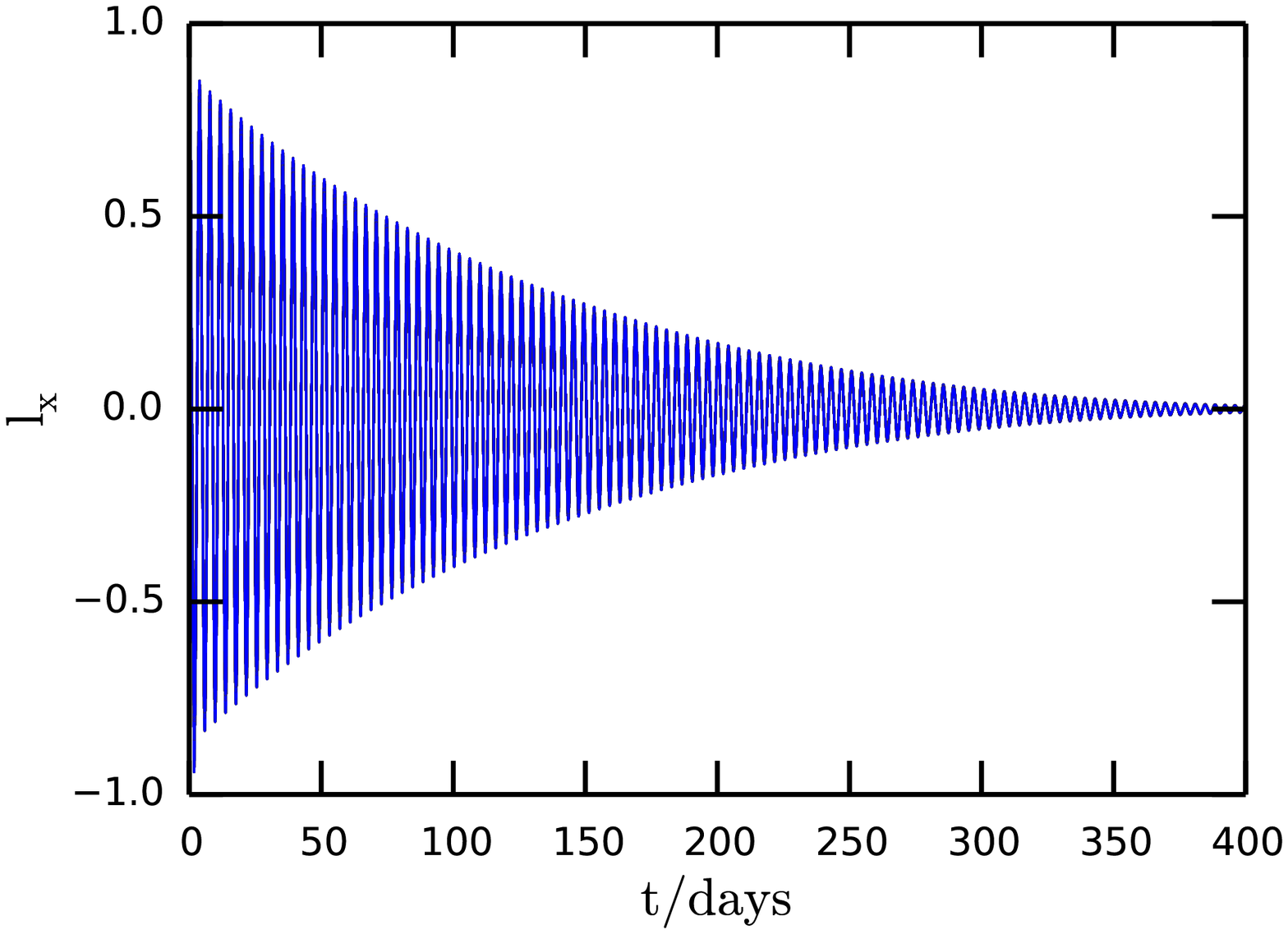}
\caption{\label{fig:prec-0.05}Evolution of the $x$-component of the disc angular momentum with time. The viscosity is $\alpha=0.05$, the value of the mass is $10^7M_{\odot}$, while the value of the spin is $a=0.2$ for the top panel and $a=0.6$ for the bottom panel.}
\end{figure}

Different values of the black hole spin correspond to different magnitude of the torque and also to different evolutions of the oscillations of the disc tilt in time, as can be seen by the comparison of the two panels of Fig. \ref{fig:prec-0.05}. Both show the evolution of the $x$-component of the disc angular momentum with time, they differ only by the spin value, which is $a=0.2$ and $a=0.6$ respectively. We note two things. Firstly, as expected, for larger $a$ the alignment is faster. Secondly, since the shape of the decay with time is different we might argue that the process that leads to alignment is not the same in the two cases.

We investigate which process is more likely to occur by computing $t_{\mathrm{align}}$ for each simulation and checking its dependence on $\alpha$.
In the cases like the one shown in the lower panel of Fig. \ref{fig:prec-0.05}, where an exponential decay with time of the $x$-component of the disc angular momentum is evident, we fitted the specific angular momentum curve with an exponential function 
\begin{equation}
l = l_{0} e^{-t/t_{\mathrm{align}}}
\label{eq:expo}
\end{equation}
and then extracted the alignment timescale.
For cases similar to the one shown in the upper panel of Fig. \ref{fig:prec-0.05}, where the shape of the decay is not exponential, rather than fitting equation (\ref{eq:expo}) to the decay function, in order to have a direct comparison with the exponential case, we have computed the alignment timescale evaluating the time after which the angular momentum value is reduced by the same factor as in the exponential case, that is by $1/e$.

\begin{figure*}
\centerline{%
\includegraphics[width=0.5\textwidth] {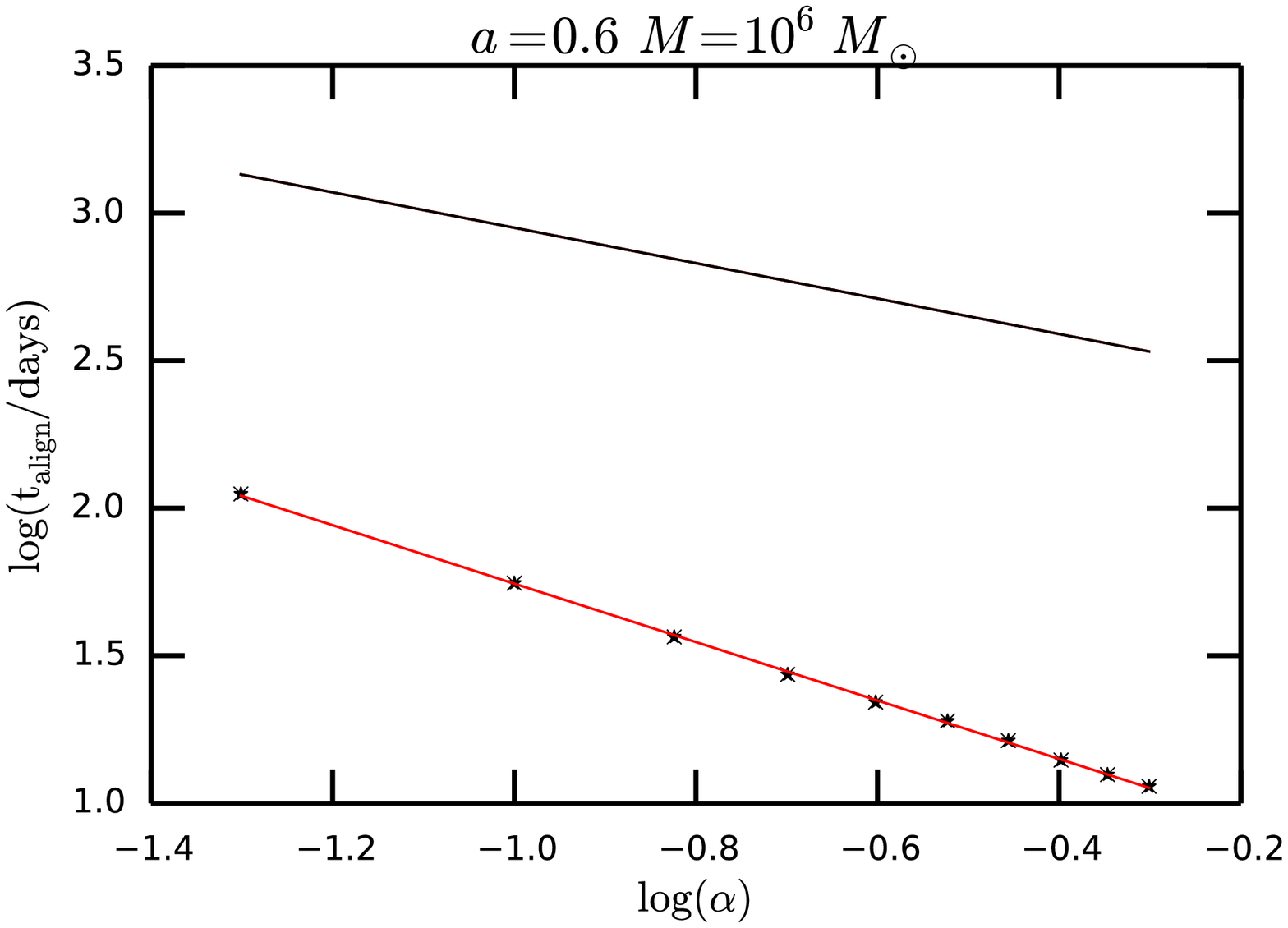}%
\includegraphics[width=0.5\textwidth] {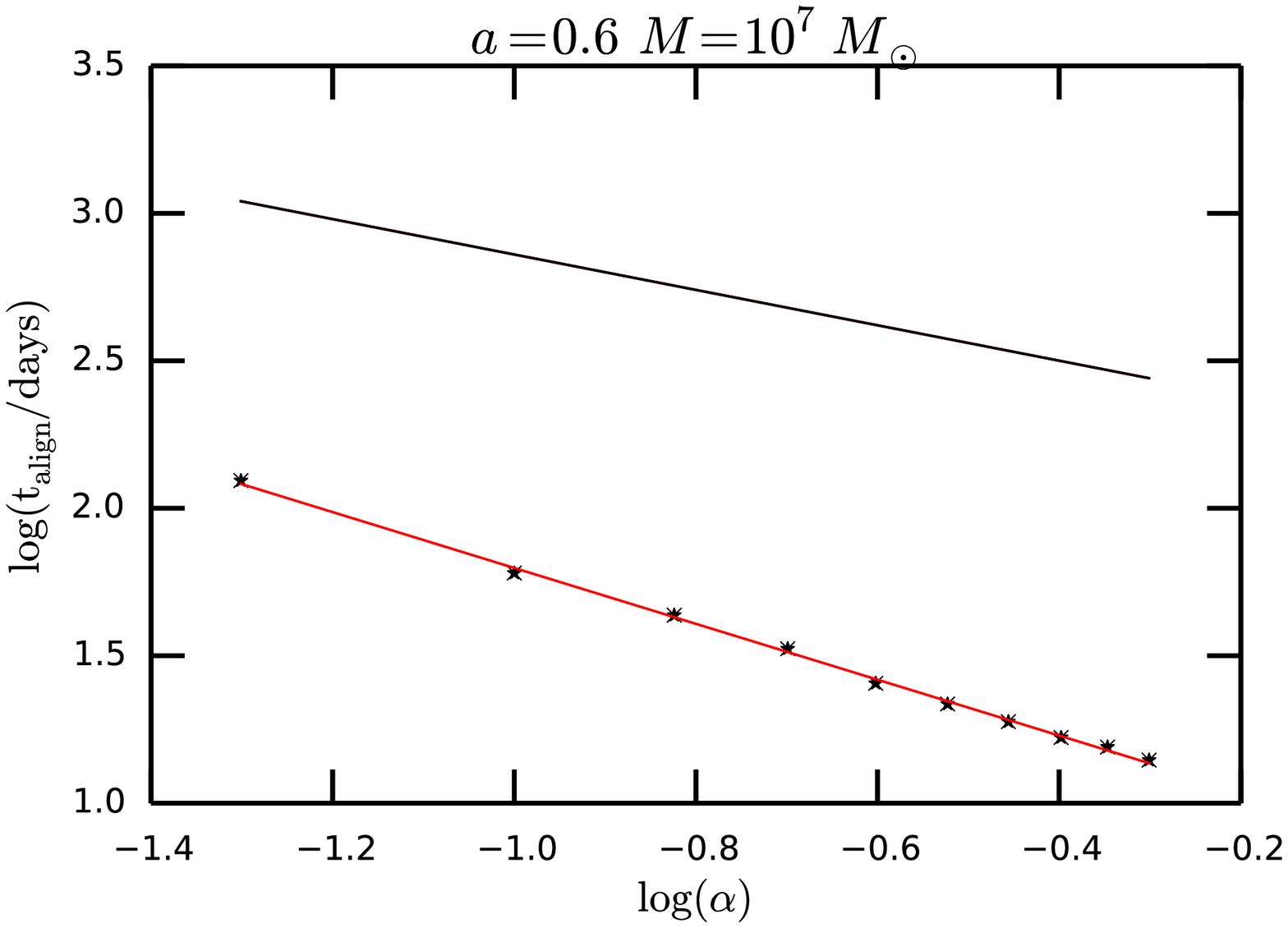}%
}%
\centerline{%
\includegraphics[width=0.5\textwidth]{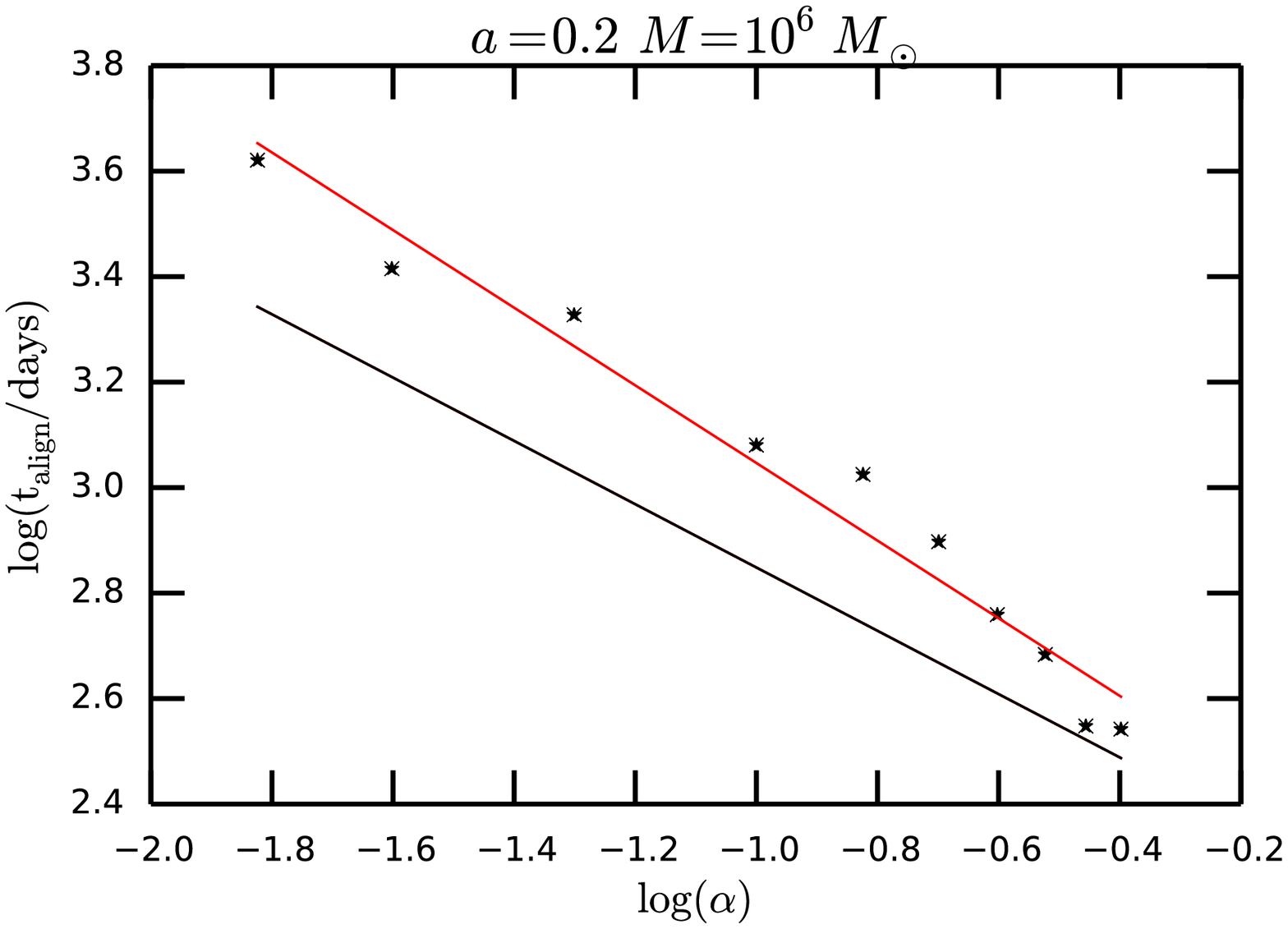}%
\includegraphics[width=0.5\textwidth]{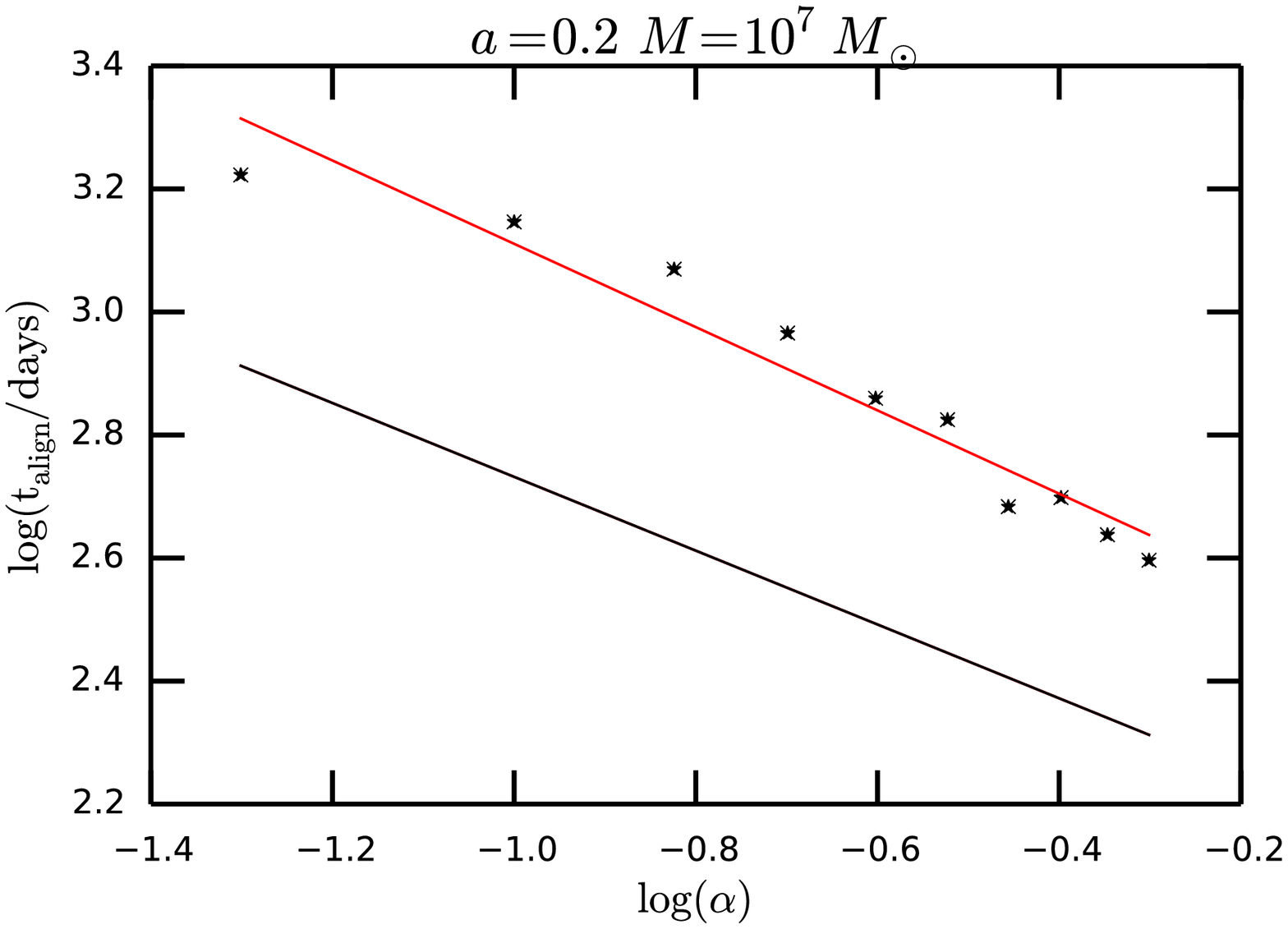}
}%
\caption{\label{fig:talign-alpha}The different panels show the alignment timescale (in days) as a function of the viscosity parameter $\alpha$ for different couples of values of the black hole parameters: mass and spin. The black and red lines are representative curves $t_{\mathrm{align}}\propto\alpha^{-s}$ with $s=0.6$ (SL) and $s=1$ respectively. The stars are the values obtained from the simulations, the points are the values computed using eq. (\ref{eq:t-thin}) predicted by SL.}
\end{figure*}

A sample of our results is shown in Fig. \ref{fig:talign-alpha}.
The top panels refer to a case of moderate spin ($a=0.6$) for the two cases $M=10^6M_{\odot}$ (left) and $M=10^7M_{\odot}$ (right). We see that the alignment time computed from the simulation is one to two orders of magnitude smaller than that predicted by SL (black line). Also, in this cases, $t_{\mathrm{align}}$ scales exactly as $\alpha^{-1}$. We conclude that viscous alignment occurs much before the disc becomes thin enough to move into the diffusive regime.

The situation becomes less clear for low spins ($a=0.2$) as shown in the bottom panel of Fig. \ref{fig:talign-alpha}. Here, the prediction by SL is much closer to the observed value. Additionally, the scaling of $t_{\mathrm{align}}$ with $\alpha$ becomes less evident (as shown by the red lines, that indicate a representative $\alpha^{-1}$ scaling).

\begin{figure}
\includegraphics[width=0.5\textwidth]{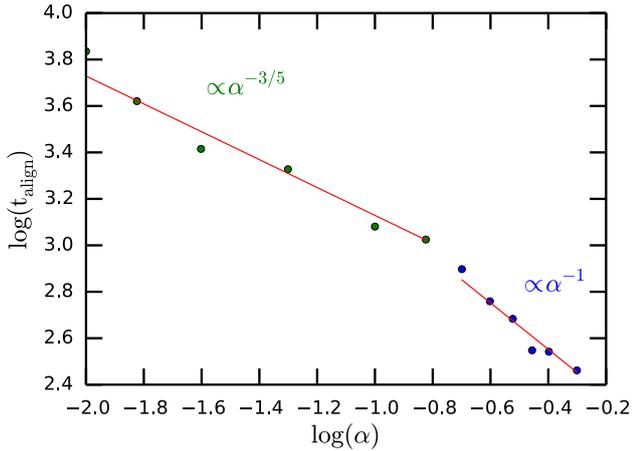}
\caption{\label{fig:fit-slopes}Alignment timescale behaviour as a function of the viscosity parameter $\alpha$ for a black hole with $a=0.2$ and $M=10^7M_{\odot}$. For low viscosity values the timescale follows the prediction of SL, while for higher values it decreases with $\alpha^{-1}$.}
\end{figure}

A closer inspection of the bottom right panel of Fig. \ref{fig:talign-alpha} reveals something interesting. While for larger values of $\alpha$ ($\alpha> 0.2$) $t_{\mathrm{align}}\propto\alpha^{-1}$, for $\alpha< 0.2$ we have $t_{\mathrm{align}}\propto\alpha^{-3/5}$, as evident in the fit shown in Fig. \ref{fig:fit-slopes}, where we have fitted $t_{\mathrm{align}}(\alpha)$ as a broken power law.

\begin{figure}
\includegraphics[width=0.5\textwidth]{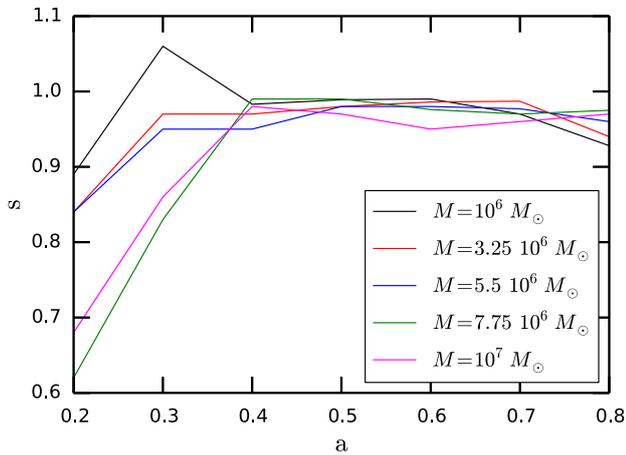}
\caption{\label{fig:slopevsM}Values of the slope parameter as a function of the black hole spin $a$ for different values of the black hole mass. For most spin values the parameter is $s=1$ almost independently on the mass.}
\end{figure}

For each value of the black hole mass and spin, we have fitted a function of the form:
\begin{equation}
t_{\mathrm{align}} = t_{0} \alpha^{-s}
\end{equation}
to the simulation data, with $t_{0}$ and $s$ as free parameters. The resulting value of $s$ as a function of $a$ for different $M$ are shown in Fig. \ref{fig:slopevsM}.

We clearly see two regimes: for $a\gtrsim 0.4$ we obtain $s=1$ independently of $M$, this indicating a viscous origin for the alignment mechanism.
For $a\leq 0.4$, $s$ is smaller than $1$. For high $M$ it is very close to $3/5$, indicating cooling as the main alignment mechanism. Values of $s$ intermediate between $3/5$ and $1$ generally indicate cases similar to the one in Fig. \ref{fig:slopevsM}, where $t_{\mathrm{align}}(\alpha)$ is a broken power-law, such that for large $\alpha$ the disc aligns viscously, while for low $\alpha$ it aligns due to cooling.

\subsection{Viscous alignment timescale calculations}\label{timescale-FL-Bate}
As shown above, the disc viscosity leads to alignment, at least for spin values $a \gtrsim 0.4$. This occurs because it acts on the shearing motions inside the disc leading to energy dissipation.
\cite{2000MNRAS.317..773B} evaluated the amount of kinetic energy dissipated by the disc viscosity and the energy dissipation rate. Then from the ratio between these two quantities the alignment timescale can be inferred. In terms of the viscous time, this is
\begin{equation}
t_{\mathrm{align}} \sim t_{\mathrm{Bate}}=\frac{1}{\alpha} \left(\frac{H}{R}\right)^2 \frac{\Omega}{\Omega_{\mathrm{p}}^{2}}\,.\label{eq:bate}
\end{equation}
This result has been obtained referring to circumbinary discs, thus for relatively thin discs with a pure power law profile of $H/R$. 

A more precise analysis has been performed by \cite{2014MNRAS.445.1731F} (FL). They argued that the damping rate of the global precession is 
\begin{equation}
\gamma = \frac{1}{t_{\mathrm{align}}}= \frac{\int_{R_{\mathrm{in}}}^{R_{\mathrm{out}}} dx\frac{4\alpha {G^2_{\phi}}}{\Sigma {c^2_{\mathrm{s}}} x^3}}{\int_{R_{\mathrm{in}}}^{R_{\mathrm{out}}} dx \Sigma x^3 \Omega}\label{eq:gamma}
\end{equation}
where $G_{\phi}$ is the internal stress in the disc given by
\begin{equation}
G_{\phi}(r) =  \int_{r_{\mathrm{in}}}^{r} dx \Sigma x^3 \Omega (\Omega_{\mathrm{p}} - Z\Omega)\,.\label{eq:gphi}
\end{equation}
where $Z$ is the nodal precession
\begin{equation}
Z(r) = \frac{\Omega^2 - \Omega^2_{\mathrm{z}}}{2\Omega^2}\,.
\end{equation}
Eq. (\ref{eq:gphi}) represents the difference between the total torque exerted by the black hole on the disc and that required in order to keep the global precession of the disc. This difference is due to the fact that the finite viscosity $\alpha$ twists the disc by an angle $\phi =\arctan (l_{y}/l_{x})$, thus creating a non-zero torque on the plane defined by the hole angular momentum and that of the disc.
Using our one-dimensional code, we evaluated the twist angle as a function of the disc radius calculated at a fixed time for each value of $\alpha$, and thus calculated the expected $t_{\mathrm{align}}$ based on FL.

\begin{figure}
\includegraphics[width=0.5\textwidth]{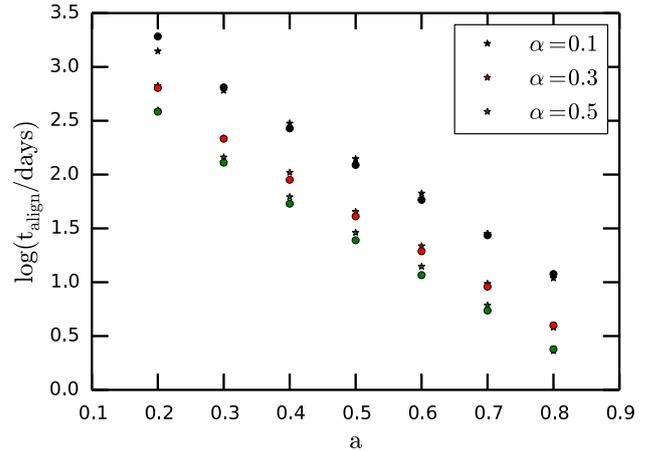}
\caption{\label{fig:focaurt-lai}Comparison between the alignment timescales (in days) obtained with the one-dimensional code and those calculated with the FL method. The latter are represented by dots while the star values are those computed with our code. We choose a black hole with $M=10^7M_{\odot}$.}
\end{figure}

The comparison between the results obtained with the FL method and those calculated with our code is shown in Fig. \ref{fig:focaurt-lai} where we can see that there is very good agreement.

This approach is equivalent to that of \cite{2000MNRAS.317..773B}, except for a shape factor $\xi$ that depends on the profiles of $G(R)$ and $\Sigma(R)$. Considering the profiles in eq. (\ref{eq:dens}) and (\ref{eq:hoverr-gen}) and assuming $\dot{m}=1$, we indeed find:
\begin{equation}
t_{\mathrm{damp}} = \frac{1}{4}\,t_{\mathrm{Bate}}\,\xi,\label{eq:gamma-new}
\end{equation}
where $\xi$ is the shape factor that lies in the range $1\leq \xi \leq 8$ and the timescale $t_{\mathrm{Bate}}$ is evaluated at the disc outer radius $R_{\mathrm{out}}$.

\section{Conclusions}\label{concl}

In this paper, we have developed a simple model for rigid disc precession in tidal disruption events. We have assumed that the disc can be modeled as a standard $\alpha$-disc dominated by radiation pressure. We have thus computed the expected precession period as a function of the main system parameters. We find periods of the order of a few days up to a few weeks, with smaller black hole spins and masses giving rise to longer periods. 

We have then compared our analytical expectations against time dependent simulations, where the disc is modeled as a series of interacting rings, applicable to small misalignments. We have thus confirmed that, initially, a TDE disc can globally precess across the whole parameter range applicable to TDE, with precessional periods matching closely the expectations of the analytical model. On a longer timescale, however, the system evolves towards alignment. For a given precessional period, a smaller black hole mass implies a faster alignment timescale, because in this case the radial extent of the disc is larger and warp communication less effective. 

Periodic modulation of a TDE signal has rarely been observed, with a tentative detection of a 2.7 d period decaying after $\approx 10$ days in the case of Swift J1644 \citep{2013ApJ...762...98L}. If we interpret this signal as arising from the Lense-Thirring precession of the disc, we would thus conclude that the black hole mass should be close to $5~10^5 - 10^6 M_{\odot}$, with a moderate spin value of $a\approx 0.6$. A smaller black hole mass would not produce such a small precession period, while a larger black hole mass would result in many more precessional cycles than observed. 

We have also investigated the mechanism responsible for alignment. SL have proposed that alignment is associated with disc cooling, which brings the disc into the diffusive warp propagation regime, where the Bardeen-Petterson effect occurs. In this case, the alignment timescale is expected to scale with viscosity as $\alpha^{-3/5}$. We have shown that such process is responsible for alignment only in a small region of the parameter space, for large black hole masses, low spins and low $\alpha$ values. For most of the cases considered, viscous dissipation of the warp is the leading alignment process, resulting in $t_{\mathrm{align}}\propto \alpha^{-1}$. 

Indeed, a theoretical calculation of the alignment timescale using the viscous dissipation model by FL is in very good agreement with our time-dependent calculations. However, the FL model is generally difficult to apply, since it requires a detailed knowledge of the disc shape and in particular of the twist, which is generally not known from observations. The simpler estimate of $t_{\mathrm{align}}$ by \citet{2000MNRAS.317..773B} (Eq. (\ref{eq:bate})), which links the alignment timescale to the disc aspect ratio, Keplerian frequency and global precession rate evaluated at the outer disc edge, provides a good first order approximation to the actual alignment time, accurate to within a factor of a few. 

Our model can be used in conjunction with future TDE observations that show an early quasi-periodicity, in order to derive the main system parameters, such as the black hole mass and especially its spin.


\section*{ACKNOWLEDGEMENTS}

We thank Chris Nixon for several useful discussions and an anonymous referee for a thorough review of the paper, which helped to significantly improve the text. SF thanks the Science and Technology Facility Council and the Isaac Newton Trust for the award of his studentship. 

\bibliographystyle{mn2e}
\bibliography{references}

\label{lastpage}

\end{document}